\documentclass[sn-mathphys,Numbered]{sn-jnl}% Math and Physical Sciences Reference Style

\usepackage{graphicx}%
\usepackage{multirow}%
\usepackage{amsmath,amssymb,amsfonts}%
\usepackage{amsthm}%
\usepackage{mathrsfs}%
\usepackage[title]{appendix}%
\usepackage{xcolor}%
\usepackage{textcomp}%
\usepackage{manyfoot}%
\usepackage{booktabs}%
\usepackage{algorithm}%
\usepackage{algorithmicx}%
\usepackage{algpseudocode}%
\usepackage{listings}%
\usepackage{siunitx}
\usepackage{float}

\begin{document}

\newcommand{\Dfrep}{$\Delta f_{\mathrm{rep}}$}
\newcommand{\frep}{$f_{\mathrm{rep}}$}
\newcommand{\ftwof}{\textit{f-2f}}
\newcommand{\dBRootHz}{\unit{\decibel}~\unit{\hertz}$^{1/2}$}
\newcommand{\dBcHz}{\unit{\decibel}c/\unit{\hertz}}

\newcommand{\nvsqrtHz}{\unit{\nano\volt}$^2$/\unit{\hertz}}
\newcommand{\subfig}[1]{\textbf{(#1)}}
\newcommand{\acetyleneFormula}{C$_2$H$_2$}
\newcommand{\ie}{\textit{i.e.}}

\newcommand{\pump}[1]{P-#1}
\newcommand{\signal}[1]{S-#1}
\newcommand{\idler}[1]{I-#1}

\renewcommand{\vec}[1]{\underline{#1}}
\newcommand{\unwrap}[1]{\mathrm{unwrap}(#1)}
\newcommand{\igmPhi}[1]{\vec{\phi}_{#1}}
\newcommand{\igmDelta}[1]{\vec{\delta}_{#1}}

\newcommand{\SNRabsolute}{SNR}
\newcommand{\SNRsqrtT}{normalized SNR}

\newcommand{\bestSNR}{50.2}
\newcommand{\bestFOM}{$3.5\times 10^8$}
\newcommand{\signalSNR}{42.3}
\newcommand{\signalSNRnorm}{40.7}
\newcommand{\DfrepNominal}{3.4~\unit{\kilo\hertz}}

\newcommand{\SignalTuningMin}{1300}
\newcommand{\SignalTuningMax}{1670}
\newcommand{\IdlerTuningMin}{2700}
\newcommand{\IdlerTuningMax}{5000}

\newcommand{\SignalTuningMinfreq}{180}
\newcommand{\SignalTuningMaxfreq}{230}
\newcommand{\IdlerTuningMinfreq}{60}
\newcommand{\IdlerTuningMaxfreq}{110}

\newcommand{\etsTime}[1]{y_{\mathrm{#1}}}
\newcommand{\etsFreq}[1]{\tilde{y}_{\mathrm{#1}}}
\newcommand{\etsSpectrum}[1]{S_{\mathrm{#1}}}
\newcommand{\absCoeff}[1]{\alpha_{\mathrm{#1}}}

\DeclareSIUnit\bar{bar}

%TC:ignore

\title[Article Title]{High-sensitivity dual-comb and cross-comb spectroscopy across the infrared using a widely-tunable and free-running optical parametric oscillator}

\author*{\fnm{Carolin P.} \sur{Bauer}}
\email{cabauer@phys.ethz.ch}
\author{\fnm{Zofia A.} \sur{Bejm}}
\author{\fnm{Michelle K.} \sur{Bollier}}
\author{\fnm{Justinas} \sur{Pupeikis}}
\author{\fnm{Benjamin} \sur{Willenberg}}
\author{\fnm{Ursula} \sur{Keller}}
\author{\fnm{Christopher R.} \sur{Phillips}}


\affil[1]{
\orgdiv{Department of Physics, Institute for Quantum Electronics}, 
\orgname{ETH Zurich}, 
\orgaddress{
%\street{Street}, \city{City}, \postcode{100190}, \state{State}, 
\country{Switzerland}}}

%\affil[2]{\orgdiv{Department}, \orgname{Organization}, \orgaddress{\street{Street}, \city{City}, \postcode{10587}, \state{State}, \country{Country}}}

\abstract{Coherent dual-comb spectroscopy (DCS) enables high-resolution measurements at high speeds without the trade-off between resolution and update rate inherent to mechanical delay scanning approaches. However, high system complexity and limited measurement sensitivity remain major challenges for DCS. Here, we address these challenges via a wavelength-tunable dual-comb optical parametric oscillator (OPO) combined with an up-conversion detection method. The OPO is tunable in the short-wave infrared (\SignalTuningMin{}-\SignalTuningMax{}~\unit{\nano\meter} range) and mid-infrared (\IdlerTuningMin{}-\IdlerTuningMax{}~\unit{\nano\meter} range) where many molecules have strong absorption bands. Both OPO pump beams are generated in a single spatially-multiplexed laser cavity, while both signal and idler beams are generated in a single spatially-multiplexed OPO cavity. The near-common path of the combs in this new configuration enables comb-line-resolved and aliasing-free measurements in free-running operation. By limiting the instantaneous idler bandwidth to below 1~\unit{\tera\hertz}, we obtain a high power per comb line in the mid-infrared of up to 160~\unit{\micro\watt}. With a novel intra-cavity nonlinear up-conversion scheme based on cross-comb spectroscopy, we leverage these power levels while overcoming the sensitivity limitations of direct mid-infrared detection, leading to a high signal-to-noise ratio (\bestSNR{}~\dBRootHz{}) and record-level dual-comb figure of merit (\bestFOM{}~\unit{\hertz}$^{1/2}$). As a proof of concept, we demonstrate the detection of methane with 2-ppm concentration over 3-\unit{\meter} path length. Our results demonstrate a new paradigm for DCS compatible with high-sensitivity and high-resolution measurements over a wide spectral range.}

\maketitle

% ############################ %

%TC:endignore
\section*{Introduction}
\label{sec:introduction}

Dual-comb spectroscopy (DCS) has advanced molecular absorption spectroscopy measurements across the infrared spectrum, including gas detection over open paths for environmental monitoring~\cite{Coburn2018, Ycas2019, Herman2023, Westberg2023, Mead2023}, combustion analysis~\cite{Schroeder2017, Makowieki2021, Yun2023}, plasma spectroscopy~\cite{Abbas2020}, and the detection of individual isotopes in gas mixtures~\cite{Muraviev2018,Parriaux2022, Egbert2024}. Many dual-comb technologies target the mid-infrared (mid-IR) wavelength region since various molecules exhibit strong and distinctive rovibrational absorption lines at these long wavelengths. However, the practicality of DCS measurements can be constrained by limited signal-to-noise ratio (SNR) and high system complexity. Here, we address both of these challenges with a new source based on a tunable optical parametric oscillator (OPO) and a novel up-conversion detection that uses the OPO as a spectroscopic transceiver to obtain sensitivity exceeding the fundamental shot-noise limited performance accessible via direct DCS detection.
%technique that greatly improves sensitivity while maintaining a simple optical system. 

Existing sources in the mid-infrared include direct laser sources such as quantum-cascade lasers~\cite{Villares2014} and inter-band cascade lasers~\cite{Sterczewski2019}, as well as parametric sources such as microresonators ~\cite{Yu2018}, difference-frequency generation~\cite{Schliesser2005,Yan2017}, optical parametric oscillators~\cite{Zhang2012,Jin2014} and intra-pulse difference-frequency generation using broadband lasers such as Cr$^{2+}$-doped ZnS/ZnSe~\cite{Bernhardt2010,Vasilyev2023, Konnov2023}. 

A critical consideration when comparing these different technologies is the spectral coverage. A broad spectral bandwidth allows for highly flexible spectroscopic measurements on a wide range of targets. Direct laser-based sources have a limited bandwidth or tuning capacity, implying many emitters are needed to cover a large fraction of the mid-IR wavelength range. In contrast, some parametric sources can cover a broad instantaneous spectral bandwidth with one device. A number of approaches in recent years have utilized dual-comb sources to access octave-spanning infrared wavelength ranges~\cite{Muraviev2018,Yu2018,Ycas2018,Timmers2018,Hoghooghi2023,Vasilyev2023, Konnov2023}. However, a broad instantaneous optical bandwidth implies that the total laser power gets distributed over a larger number of comb lines, which impairs the achievable sensitivity~\cite{Newbury2010}; we follow the approach of \cite{Newbury2010} where the \SNRabsolute{} is proportional to the magnitude of the Fourier transform of the time-domain interferogram. So far, for high-resolution octave-spanning DCS experiments in the mid-IR, this led to either integration times in the minute-to-hour time scale or an \SNRabsolute{} limited to $<$20~\dBRootHz{}~\cite{Muraviev2018,Vasilyev2023,Ycas2018,Timmers2018,Hoghooghi2023}. One method to decrease the number of comb lines is to operate at high repetition rates, but this comes with two challenges: It decreases the available pulse energy for nonlinear wavelength conversion processes, and for high-resolution measurements it becomes necessary to carefully interleave the DCS measurement by sweeping the comb line frequencies~\cite{Villares2014,Komagata2023}. 

\begin{figure}[H] % [h]%
\centering
\includegraphics[width=0.9\textwidth]{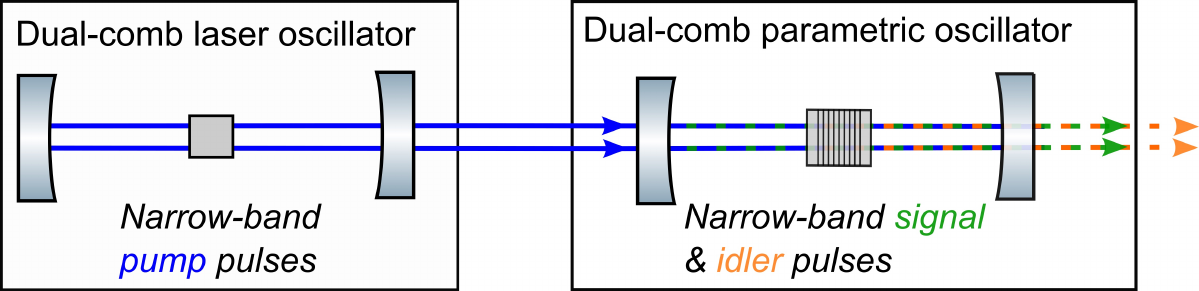}
\caption{
Block diagram of the system used for dual-comb spectroscopy. Single-cavity narrow-band dual-comb laser oscillator synchronously pumps a wavelength-tunable single-cavity dual-comb optical parametric oscillator.
}
\label{fig:system_concept}
\end{figure}

A promising alternative approach is a dual-comb source with a narrow instantaneous bandwidth where the broad spectral coverage is obtained via wavelength tunability. In this configuration, there are only a few thousand comb lines in the optical spectrum at a time, in contrast to octave-spanning 100-\unit{\mega\hertz} lasers with of order a million comb lines. Therefore, it supports high power per comb line, and \unit{\kilo\hertz}-level repetition rate differences can be used without aliasing of the DCS spectrum. The large repetition rate difference enables phase tracking required for comb-resolved (and hence fully coherent) measurements~\cite{Hebert2019} with simple free-running dual-comb oscillators~\cite{Roy2012,Hebert2017,Camenzind2023,Phillips2023}. To realize such a spectrometer configuration, here we demonstrate for the first time a spatially-multiplexed single-cavity dual-comb OPO synchronously pumped by a spatially-multiplexed single-cavity dual-comb laser. We thereby obtain a simple yet powerful dual-comb source emitting in the SWIR and mid-IR using two free-running cavities (one laser cavity, and one OPO cavity) enabling high-sensitivity measurements. The concept of the light source is illustrated in Fig.~\ref{fig:system_concept}.	

Singly-resonant OPOs are ideal for obtaining tunable coherent light with narrow instantaneous bandwidth. While leveraging the high conversion efficiency of parametric processes, they transfer the low-noise and high powers achievable with near-infrared bulk-crystal solid-state pump sources~\cite{Benedick2012,Endo2018} to the SWIR and mid-IR over a broad spectral range without having to rely on spectral broadening~\cite{Burneika1972,Edelstein1989,Dudley1994}. To take advantage of these capabilities, we have developed a new Yb:YAG dual-comb pump laser oscillator which operates at a repetition rate of 250~\unit{\mega\hertz} and delivers $\approx 2~\unit{\watt}$ average power per comb, which enables several hundred \unit{\milli\watt} of signal and idler power. The moderate comb-line spacing is able to resolve the absorption features of various gases at low pressure (e.g. $<$100~\unit{\milli\bar}), and the combination of picosecond pulse durations and hundreds of \unit{\milli\watt} average powers yields more than 100~\unit{\micro\watt} per comb line, compatible with high-sensitivity measurements. 

While tunable OPOs offer high performance, a major concern is system complexity since potentially four cavities are needed with corresponding feedback loops and matched tuning servos to keep them mutually stabilized and spectrally overlapped. Here we address this complexity issue by using single-cavity dual comb concepts, a topic which has been discussed in recent reviews~\cite{Liao2020, Zhang2023}. In the spatial multiplexing approach of \cite{Pupeikis2022}, one of the flat intra-cavity mirrors is replaced with a reflective biprism in order to create two closed cavity paths (one for each comb). Tuning the lateral biprism position adjusts the cavity round-trip time difference between the two combs. Here we use this technique for both the laser and OPO cavities simultaneously.

A variety of shared-cavity approaches have been explored for dual-comb OPOs. There are demonstrations based on two pump lasers pumping a single OPO cavity with either two nonlinear crystals~\cite{Abbas2019} or asynchronously with one nonlinear crystal~\cite{Zhang2013,Jin2015}. These systems are partly stabilized and had limitations to their sensitivity related either to decoherence over short time scales or to low power per comb line~\cite{Zhang2013,Jin2015}. A notable recent demonstration showed a cw-OPO cavity pumped by an electro-optic dual-comb source~\cite{Long2024}. In previous work~\cite{Bauer2022}, we demonstrated a simple free-running single-cavity approach for both the pump laser and the OPO. In the OPO ring cavity with counter-propagating beams, the intra-cavity dispersion led to a compensation of the optical path length difference for the two combs via a shift of the center wavelength. However, this reduced spectral overlap between the two combs limited the repetition rate difference \Dfrep{} to the 100-\unit{\hertz} range. 

In this work, by using spatial multiplexing of a linear OPO cavity, we can freely adjust the optical path length difference between the two resonant signal beams in order to match their round-trip times with the inverse of the repetition rate of their respective pump pulse trains. Accordingly, the synchronous pumping condition can be satisfied for both combs simultaneously at repetition rate differences \Dfrep{} of several kilohertz, allowing high-resolution and high-sensitivity coherently-averaged measurements. The OPO is tunable over a wavelength range \SignalTuningMin{}~\unit{\nano\meter} to \SignalTuningMax{}~\unit{\nano\meter} (signal) and \IdlerTuningMin{}~\unit{\nano\meter} to \IdlerTuningMax{}~\unit{\nano\meter} (idler), with up to 580~\unit{\milli\watt} (signal) and 320~\unit{\milli\watt} (idler). To validate that the configuration is compatible with coherent DCS, we conduct spectroscopy measurements using the OPO signal combs around 1540~\unit{\nano\meter}. Our results show good agreement with the HITRAN database~\cite{Kochanov2016} and a high \SNRabsolute{} of \signalSNR{}~\unit{\decibel} within a measurement time of 2.15~\unit{\second} (figure of merit of $3.2\times10^7$~\unit{\hertz}$^{1/2}$~\cite{Guay2023}). 

For practical spectroscopy measurements in the mid-IR, photodetection poses an additional challenge. Mid-IR detectors have limited saturation power, and typically exhibit lower quantum efficiency, lower speed, higher cost, and higher noise compared to InGaAs detectors. Our new dual-comb OPO architecture enables an elegant solution to this issue by using nonlinear up-conversion to operate the OPO cavity as a spectroscopic dual-comb transceiver. To accomplish this, we mix the idler of one comb with the intra-cavity signal of the other comb, thereby efficiently up-convert the mid-IR spectroscopic information (encoded in the free induction decay (FID) of the idler pulses) to the near-IR in a time-gated fashion. Now the up-converted wave can be detected with a standard InGaAs detector in a heterodyne scheme. Our approach builds on the recently demonstrated cross-comb spectroscopy (CCS) technique \cite{Liu2023} with two new and critical milestones: use of a tunable dual-comb OPO (which also removes the need for any spectral broadening stage), and the development of an efficient and experimentally simple intra-cavity up-conversion method.

Rather than having to attenuate the mid-IR beam to avoid detector saturation, we use the full $>$300-\unit{\milli\watt} average power of the idler combs in this process. Due to this fundamental boost in sensitivity compared to direct dual-comb detection, we can obtain a high \SNRsqrtT{} of \bestSNR{}~\dBRootHz{} in a proof-of-principle measurement of ambient air at 3320~\unit{\nano\meter}. Also for the corresponding figure of merit (FOM)~\cite{Connes1971, Newbury2010, Guay2023} a high value of \bestFOM{}~\unit{\hertz}$^{1/2}$ is achieved, which is well above the fundamental quantum-noise-limited value that could be obtained by DCS using the same local oscillator power~\cite{Newbury2010}. The high sensitivity enables the trace-gas detection of methane at its natural abundance of about 2~ppm over a path length of 3~\unit{\meter}. We conduct two further CCS measurements across the OPO wavelength tuning range on a low-pressure acetylene sample cell to demonstrate the broad spectral coverage and high resolution that can be obtained with this system. These results represent, to the best of our knowledge, the highest SNR and the highest FOM obtained to date in high-resolution mid-IR DCS or CCS in a traveling-wave configuration~\cite{Muraviev2018, Ycas2018, Chen2020, Komagata2023,Vasilyev2023, Liu2023, Konnov2023}.

\section*{Results}

\subsection*{Experimental set-up and operating conditions}
\label{sec:results_setup}

The laser source consists of a dual-comb diode-pumped solid state laser (SSL) from a single cavity pumping a dual-comb OPO from a single cavity (Fig.~\ref{fig:system_concept}). Both cavities are spatially multiplexed with an intra-cavity biprism that introduces a spatial separation on the optically active elements while still sharing all optical elements. Since we operate the cavities in free-running mode, it is necessary to measure the phase fluctuations fast enough in order to track them. There is a natural phase-sampling rate of \Dfrep{} because a single carrier-envelope phase value can be obtained from each dual-comb interferogram. Therefore, to conduct coherent measurements without the aid of cw lasers~\cite{Deschenes2010}, \Dfrep{} should be sufficiently high to sample the free-running fluctuations of the dual-comb laser. The shared cavity approach leads to correlated fluctuations of the combs, which enables phase tracking at smaller \Dfrep{}. This is beneficial since \Dfrep{} also needs to be small enough to avoid spectral aliasing. The RF bandwidth occupied by the interferograms is
\begin{align}\label{eq:RFbandwidth}
    \Delta f_{RF} = \Delta\nu_{opt}\frac{\Delta f_{\mathrm{rep}}}{f_{\mathrm{rep}}}
\end{align}
where $\Delta\nu_{\mathrm{opt}}$ is the full optical bandwidth of the comb. To avoid aliasing, this RF bandwidth must not exceed $f_{\mathrm{rep}}/2$. The condition is fulfilled with an optical bandwidth constrained according to:
\begin{align}
    \Delta\nu_{\mathrm{opt,max}}=\frac{f_{\mathrm{rep}}^2}{2}\frac{1}{\Delta f_{\mathrm{rep}}}
\end{align}
Therefore, high \Dfrep{} requires either increasing \frep{} or narrowing the optical bandwidth. Here, we choose a configuration with moderate \frep{}=250~\unit{\mega\hertz} and picosecond pulse durations for several reasons: (i) this provides high spectral resolution without comb line interleaving, (ii) optical pulse parameters well-suited for efficient frequency conversion, and (iii) high power per comb line enabling high-sensitivity. Consequently, Yb:YAG is selected as a gain medium for the OPO pump since its modest gain bandwidth naturally leads to transform-limited pulses of several hundred femtoseconds duration. When at the same time the tuning potential of an OPO is exploited, the spectrum can be positioned at a specific set of absorption lines to efficiently capture the spectroscopic signatures of various molecules over a broad wavelength range.

\subsection*{Spatially-multiplexed dual-comb Yb:YAG laser}
\label{sec:results_YAG}

\begin{figure}[H] % [h]%
\centering
\includegraphics[width=0.9\textwidth]{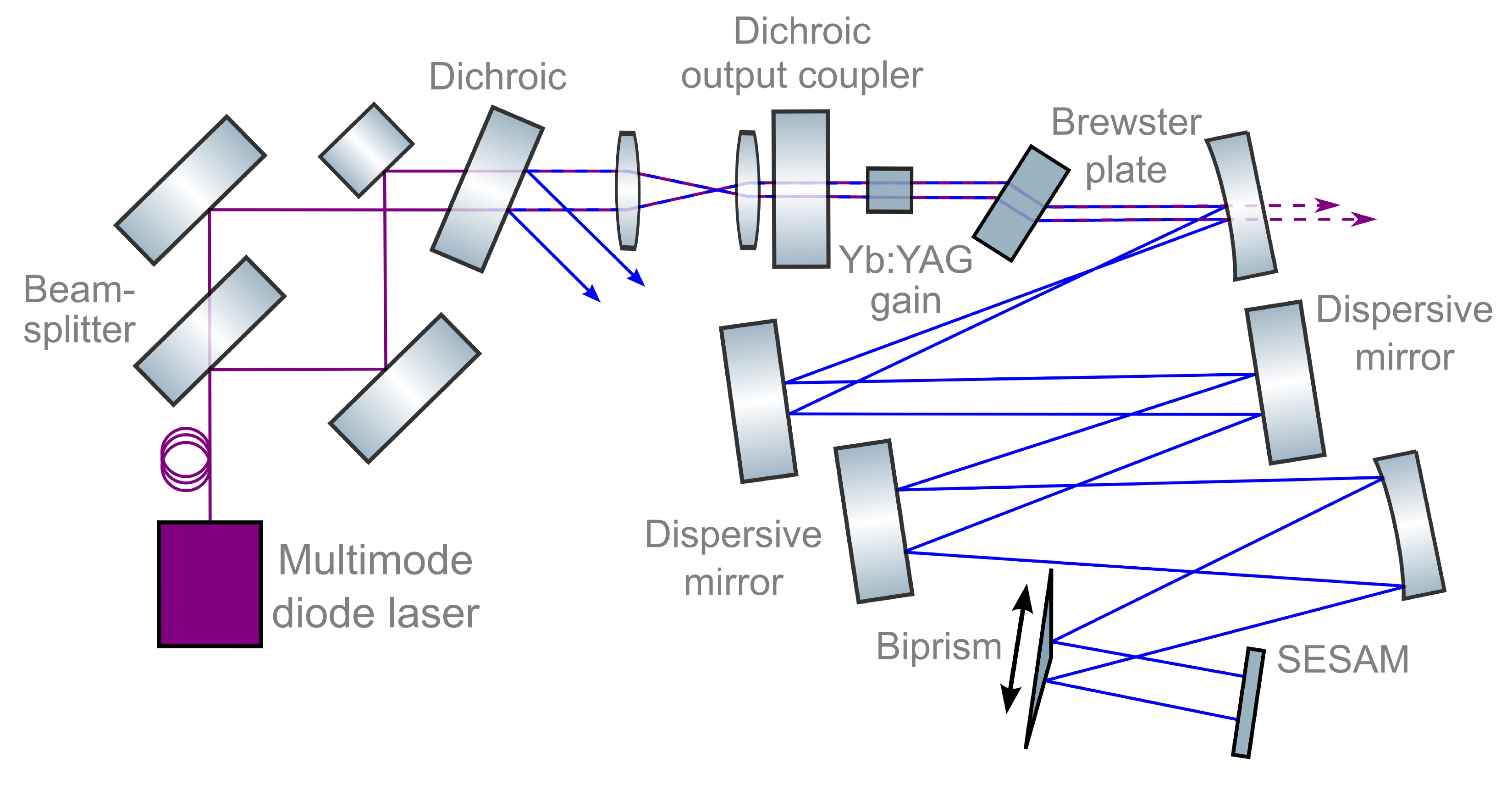}
\caption{
Schematic of the dual-comb Yb:YAG laser and its pumping arrangement. The laser is end-pumped by a multi-mode diode laser delivering 18.7~\unit{\watt} at 969~\unit{\nano\meter} (DILAS). The pump mirror is dichroic for the diode pump light and the solid-state laser light and serves as the output coupler (OC) with a transmission rate of 3.45\%. The combination of an undoped YAG Brewster plate introducing self-phase modulation and dispersive mirrors introducing an estimated round-trip group delay dispersion (GDD) of -13000~\unit{\femto\second}$^2$ provides the conditions for stable soliton modelocking. The biprism is mounted on a translation stage to allow for tuning of the repetition rate difference \Dfrep{}.
}
\label{fig:YAG_laser}
\end{figure}

\begin{figure}[H] % [h]%
\centering
\includegraphics[width=0.9\textwidth]{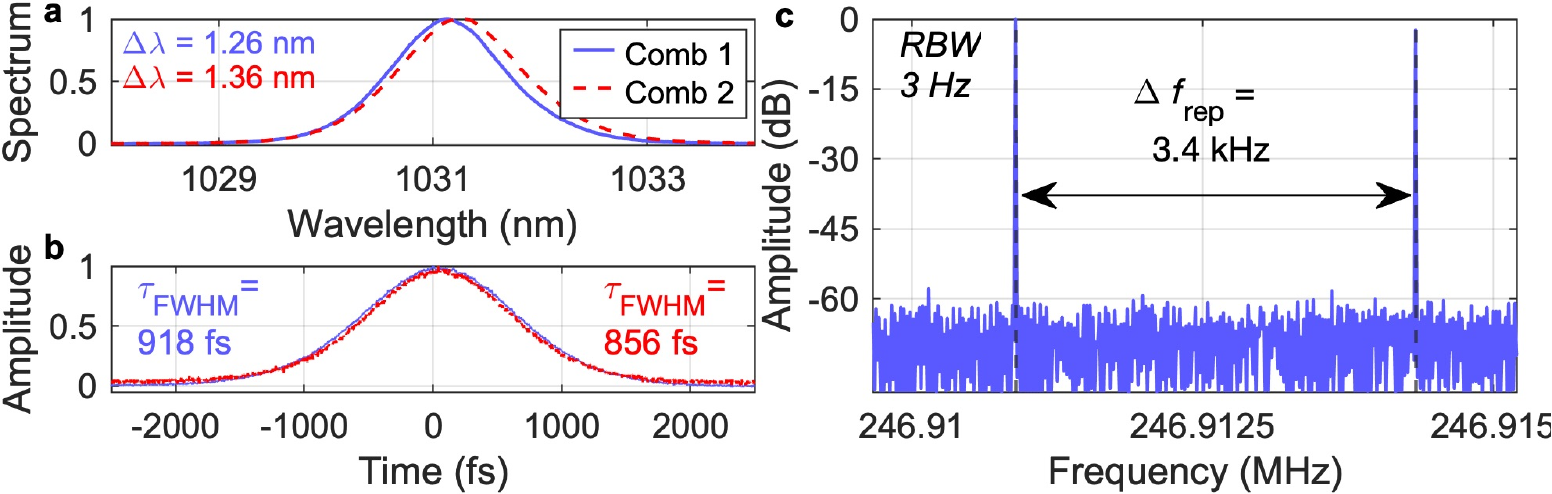}
\caption{
Characterization of the single-cavity dual-comb Yb:YAG pump laser. \\
\subfig{a} Optical spectrum at a center wavelength of 1031~\unit{\nano\meter} for comb 1 (blue) and comb 2 (red); resolution bandwidth (RBW): 0.08~\unit{\nano\meter}, measured with a diffraction grating-based spectrum analyzer (Keysight-Agilent 70951A). \\
\subfig{b} Second-harmonic generation autocorrelation traces with corresponding deconvoluted full-width half-maximum (FWHM) pulse durations of 920~\unit{\femto\second} (comb 1) and 860~\unit{\femto\second} (comb 2). \\
\subfig{c} RF spectrum of the pump laser. The nominal repetition rate difference is \Dfrep{}=\DfrepNominal{} at a repetition rate of 246.9~\unit{\mega\hertz}, where comb 1 is defined to be at the lower repetition rate; RBW: 3~\unit{\hertz}. 
}
\label{fig:YAG_performance}
\end{figure}

The Yb:YAG pump laser (see Fig.~\ref{fig:YAG_laser}) generates transform-limited picosecond pulses via soliton modelocking with a semiconductor saturable absorber mirror (SESAM)~\cite{Keller1992,Keller1996}. A dual-comb oscillator configuration is realized by inserting a biprism-like optic with a high-reflectivity coating as a turning mirror into the cavity~\cite{Pupeikis2022}. The biprism apex angle of 179\unit{\degree} leads to a horizontal mode center separation of 0.7~\unit{\milli\meter} on the gain medium and 2.0~\unit{\milli\meter} on the biprism and the SESAM. Lateral translation of the biprism changes the cavity length difference between the two combs, resulting in a tuning range of \Dfrep{} from -100~\unit{\hertz} to 3.9~\unit{\kilo\hertz} without impairing the laser performance. The laser uses a single multimode-fiber-coupled pump diode that is split into two beams of equal power to pump each comb.  The laser is implemented in a robust aluminum housing to reduce its susceptibility to environmental fluctuations. For comparable lasers, a stability of \Dfrep{} in the ~\unit{\micro\hertz} range was obtained with the help of a slow piezo-based feedback loop to correct for small long-term drifts in \Dfrep{} that were measured with a frequency counter approach~\cite{Pupeikis2022}. Figure~\ref{fig:YAG_performance} shows the performance of the laser at a nominal operation point at which it pumps the OPO. The average output powers are 2.0~\unit{\watt} (comb 1) and 1.9~\unit{\watt} (comb 2), which corresponds to optical-to-optical efficiencies of 23\% (comb 1) and 22\% (comb 2). 

\subsection*{Spatially-multiplexed dual-comb OPO}
\label{sec:results_OPO_cavity}

\begin{figure}[H] % [h]%
\centering
\includegraphics[width=0.9\textwidth]{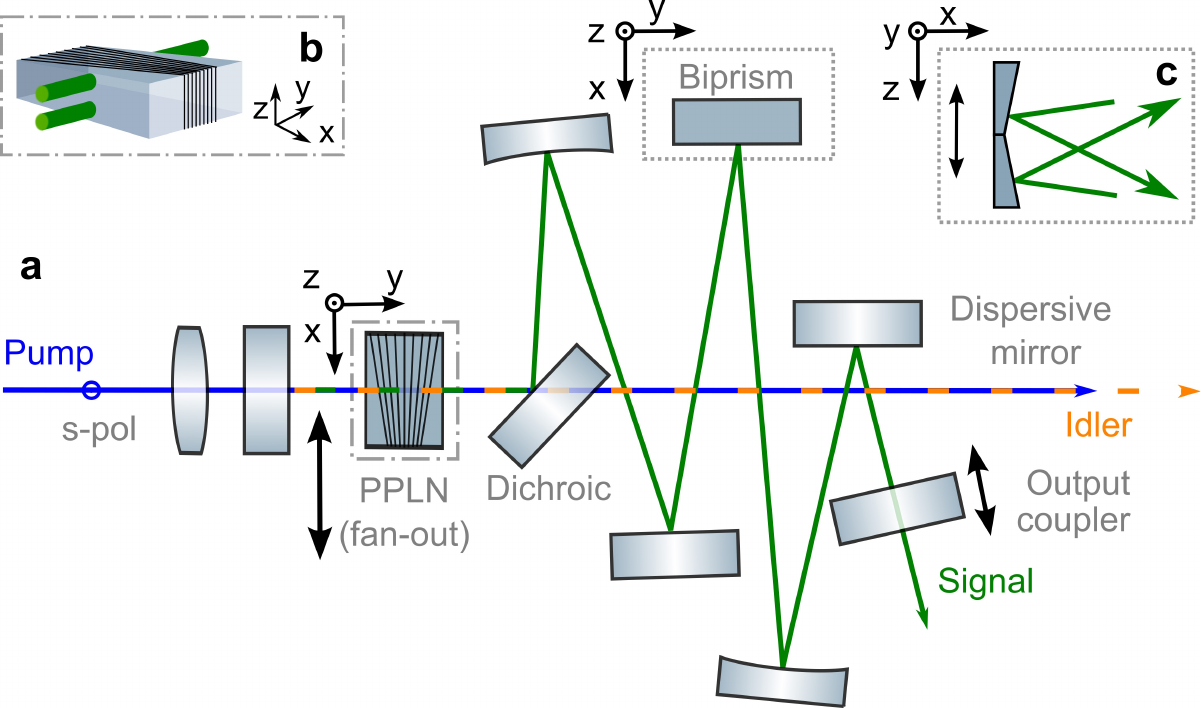}
\caption{
\subfig{a} Schematic of the dual-comb OPO. \\
\subfig{b} Illustration of the spatial multiplexing in the PPLN crystal, where the two combs are spatially separated in the direction vertical relative to the optical table and interact with the same QPM grating period. \\
\subfig{c} Illustration of the spatial multiplexing on the biprism in the direction vertical relative to the optical table.
}
\label{fig:OPO_schematic}
\end{figure}

The OPO is synchronously and collinearly pumped by the two output beams of the Yb:YAG laser. The OPO cavity layout is depicted in Fig.~\ref{fig:OPO_schematic}a. Analogous to the SSL, the OPO is spatially multiplexed with a biprism (see Fig.~\ref{fig:OPO_schematic}c). This allows to implement a linear cavity that is singly resonant for both signal beams. The beams are spatially separated in the nonlinear crystal (see Fig.~\ref{fig:OPO_schematic}b), which avoids nonlinear coupling between the two combs. The round-trip time difference between the two signal combs in the OPO cavity is modified by tuning the biprism position transverse to the beam propagation direction. Thus, the round-trip time of each signal comb can be matched to the inverse repetition rate of each pump pulse train. By this means, both combs can be synchronously pumped simultaneously without requiring a shift of the center wavelength to compensate for the relative round trip delay from \Dfrep{}. 

The OPO is pumped with 1.7~\unit{\watt} (comb 1) and 1.6~\unit{\watt} (comb 2) in the periodically-poled LiNbO$_3$ (PPLN) crystal. Both combs have the same polarization to utilize the $\mathrm{d}_{33}$ nonlinear coefficient. The measured pump 1/e$^2$ radius at the position of the crystal is 60~\unit{\micro\meter} and the calculated mode radius for the signal beam is 80~\unit{\micro\meter}. The idler beam is transmitted by a 45\unit{\degree} dichroic designed for s-polarization: it has transmission $T>70$\% from 2500-5000~\unit{\nano\meter} and reflectivity $R>$99.85\% from 1290-1650~\unit{\nano\meter}. The s-polarization requires the orientation of the PPLN crystal as illustrated in Fig.~\ref{fig:OPO_schematic}b. To ensure both combs experience the same quasi-phase-matching (QPM) grating period, the multiplexing of the OPO cavity is implemented in the vertical, see Fig.~\ref{fig:OPO_schematic}b-c. The 4-\unit{\milli\meter}-long crystal, 5~\unit{\mol}.~\% MgO doped, has a QPM fan-out grating with periods ranging from 24.8~\unit{\micro\meter} to 31.2~\unit{\micro\meter} (HC Photonics) and is operated at room temperature. The crystal is anti-reflection (AR) coated on both facets with specifications $R<1\%$ for 1020-1040~\unit{\nano\meter} and 1290-1620~\unit{\nano\meter}, and $R<10\%$ for 3000-5000~\unit{\nano\meter}. In order to optimize the oscillation threshold to obtain high pump depletion, we use a wedged OC with an out-coupling rate $T_\mathrm{OC}=12$\% (Layertec, 1250-1700~\unit{\nano\meter}). At the nominal operation point, the OPO runs 2.7x (comb 1) and 2.8x (comb 2) above the oscillation threshold. The flat OPO end mirror is mounted on a motorized picomotor translation stage to adjust the average cavity round-trip time. Synchronous pumping for both combs simultaneously is obtained by adjusting the end mirror and biprism positions. 

A dispersive mirror adding -800~\unit{\femto\second}$^2$ per round-trip (Ultrafast Innovations, R $>99.8$\%, 1325-1725~\unit{\nano\meter}) is included in the cavity. The increased magnitude of the round-trip dispersion decreases the sensitivity of the center wavelength to cavity length fluctuations. In a measurement where we tune the center wavelength over the full tuning range and determine the required cavity length change to maintain synchronicity, we estimate a total round-trip intra-cavity GDD of -600~\unit{\femto\second}$^2$. 

The OPO cavity was optimized to support a small spatial separation between the two signal combs as this leads to more correlated path length fluctuations and hence low noise, thereby permitting operation at small \Dfrep{}. The cavity design, which maintains a small vertical separation between the two combs below 1.6~\unit{\milli\meter} throughout the cavity, required an inverted biprism-like optic that reflects the incoming beams towards a common optical axis (seeFig.~\ref{fig:OPO_schematic}c); the required optic was assembled by gluing two separate angled pieces together. 
 
\subsection*{Characterization of the dual-comb system}
\label{sec:results_OPO_output}

In this section, we present the performance characteristics of the dual-comb system. For the relative intensity noise (RIN) and SWIR spectroscopy measurements, the OPO is tuned to about 1545~\unit{\nano\meter}. The results of the OPO characterization at this tuning point are summarized in Table~\ref{tab:OPO}.

\newcommand{\RINTableRow}{RIN at $f_{\mathrm{offset}}>1$~\unit{\mega\hertz}}
\begin{table}[h]
\caption{Summary of the characterization of the OPO output beams.}
\label{tab:OPO}
\begin{tabular*}{\textwidth}{@{\extracolsep\fill}lcccccc}
%\begin{tabular}{@{}llll@{}}
    \toprule%
    Parameter                           & Unit               & Signal 1 & Signal 2 & Idler 1 & Idler 2 \\
    \midrule
    Center wavelength, $\lambda_0$      & \unit{\nano\meter} & 1546.8     & 1546.7     & 3091    & 3093 \\
    Bandwidth, $\Delta\lambda$ (FWHM)   & \unit{\nano\meter} & 5.4      & 6.2      & 20.6    & 21.9 \\
    Bandwidth, $\Delta\nu$ (FWHM)       & \unit{\tera\hertz} & 0.68     & 0.78     & 0.65    & 0.69 \\
    $P_{\mathrm{out}}$                  & \unit{\milli\watt} & 580      & 585      & 320     & 322 \\
    Power per comb line                 & \unit{\micro\watt} & 212       & 186       & 122     & 116 \\
    \RINTableRow{}                      & \dBcHz{}           & -162.8     & -160.1     & -       & - \\
    \botrule
\end{tabular*}
%\end{tabular}
%
%\footnotetext{...}
%\footnotetext[1]{...}
\end{table}
 
To determine the quantum efficiency of the parametric amplification process, we account for transmission losses experienced by the two beams on the way to the power meter to estimate the average pump and idler powers inside the PPLN crystal. This estimation indicates internal idler powers of 366~\unit{\milli\watt} (for comb 1) and 369~\unit{\milli\watt} (for comb 2), yielding quantum efficiencies of $>$60\%.

Another important performance aspect for high-sensitivity DCS measurements is RIN. SSLs achieve ultra-low noise performance at the high offset frequencies relevant for DCS measurements due to their powerful low-loss cavities and long upper state lifetime gain media \cite{Shoji2016}. OPOs can inherit this property from their pump laser~\cite{Burneika1972,Edelstein1989,Dudley1994}, making them a compelling choice for low-noise tunable sources. The noise characterization was performed on the signal beams (see supplementary information). For both the pump laser and the OPO the measured RIN is lower than –160~\dBcHz{} for offset frequencies beyond 1~\unit{\mega\hertz}, which corresponds to shot-noise limited performance given the illumination levels on the photodiode.

\begin{figure}[H] % [h]%
\centering
\includegraphics[width=0.9\textwidth]{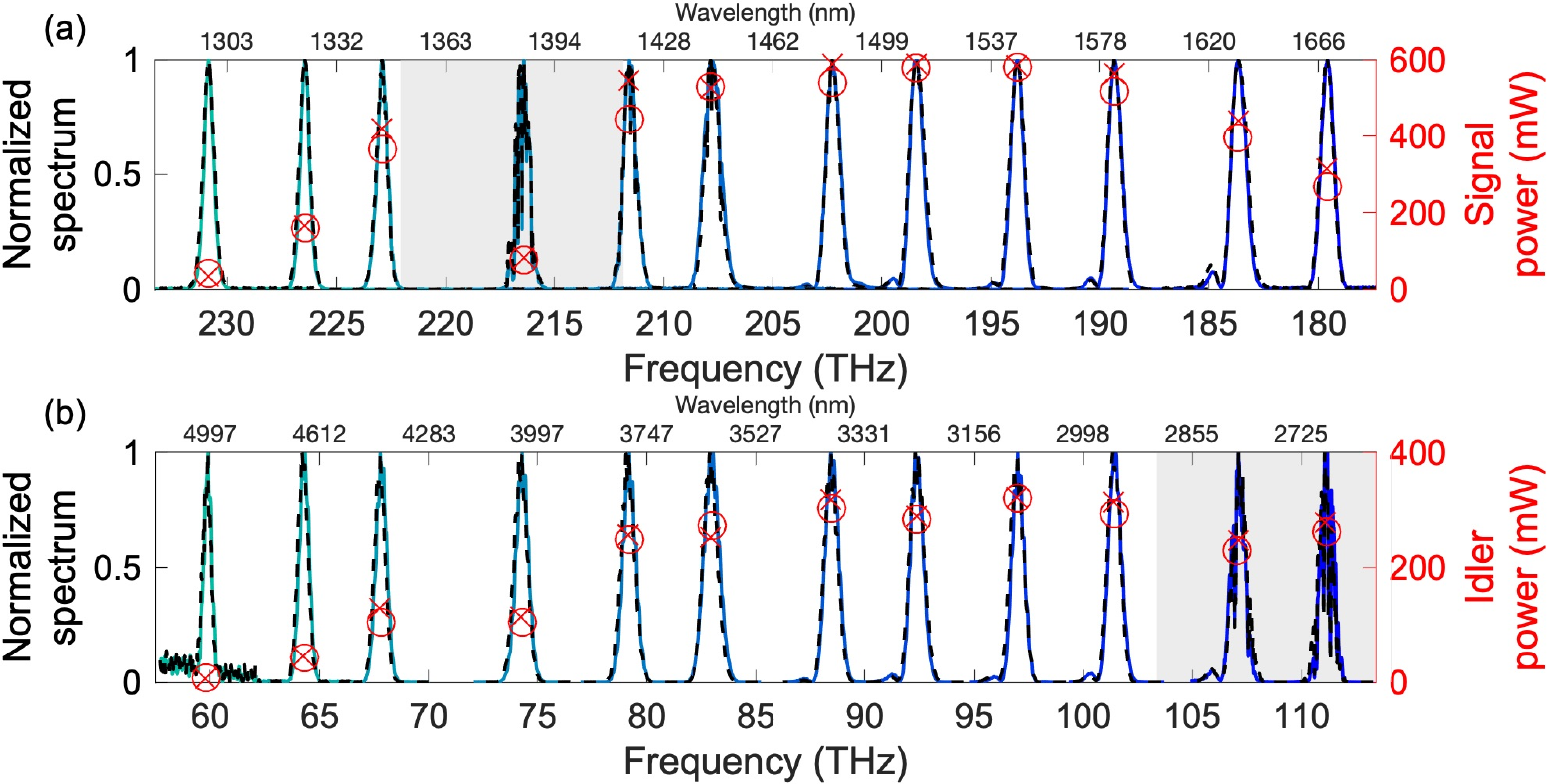}
\caption{
Wavelength tunability of the OPO. \\
\subfig{a} Optical spectra of signal 1 (colored lines) and signal 2 (black dashed lines). RBW: 0.08~\unit{\nano\meter}, measured with a diffraction grating-based spectrum analyzer (Keysight-Agilent 70951A). The corresponding power levels for signal 1 (circles) and for signal 2 (crosses). \\
\subfig{b} Optical spectra of idler 1 (colored lines) and idler 2 (black dashed lines). The corresponding power levels for idler 1 (circles) and for idler 2 (crosses). The idler power is calculated to the position of the idler transmitting dichroic by considering the losses from a coated Germanium window, an uncoated CaF$_2$ lens, and three silver mirrors. RBW: 0.09~\unit{\nano\meter}, measured with a Fourier transform-based spectrum analyzer (Thorlabs OSA205). The spectra of the same color correspond to a signal-idler pair generated at a distinct phase-matching configuration. The shaded areas indicate the water absorption windows in the signal and idler tuning range.
}
\label{fig:OPO_tuning}
\end{figure}

Figure~\ref{fig:OPO_tuning} displays the tuning range of the OPO accessed by scanning the lateral position of the fan-out PPLN crystal. The tuning range for the idler reaches from \IdlerTuningMin{}~\unit{\nano\meter} to \IdlerTuningMax{}~\unit{\nano\meter} (\IdlerTuningMinfreq{}-\IdlerTuningMaxfreq{}~\unit{\tera\hertz}) and for the signal from \SignalTuningMin{}~\unit{\nano\meter} to \SignalTuningMax{}~\unit{\nano\meter} (\SignalTuningMinfreq{}-\SignalTuningMaxfreq{}~\unit{\tera\hertz}). Due to the dispersion in the cavity, its length is adjusted at each tuning step to maintain synchronicity with the pump. The output power exhibits some variation with respect to wavelength, which is likely due to a combination of effects including wavelength dependence of the coatings (for  $\lambda_s<1350~\unit{\nano\meter}$ and $\lambda_s>1600~\unit{\nano\meter}$), water absorption (for $\lambda_s$ from 1350-1415~\unit{\nano\meter}), and the parametric gain.

\subsection*{SWIR dual-comb spectroscopy of acetylene} 
\label{sec:results_DCS_SWIR}

\begin{figure}[H] % [h]%
\centering
\includegraphics[width=0.9\textwidth]{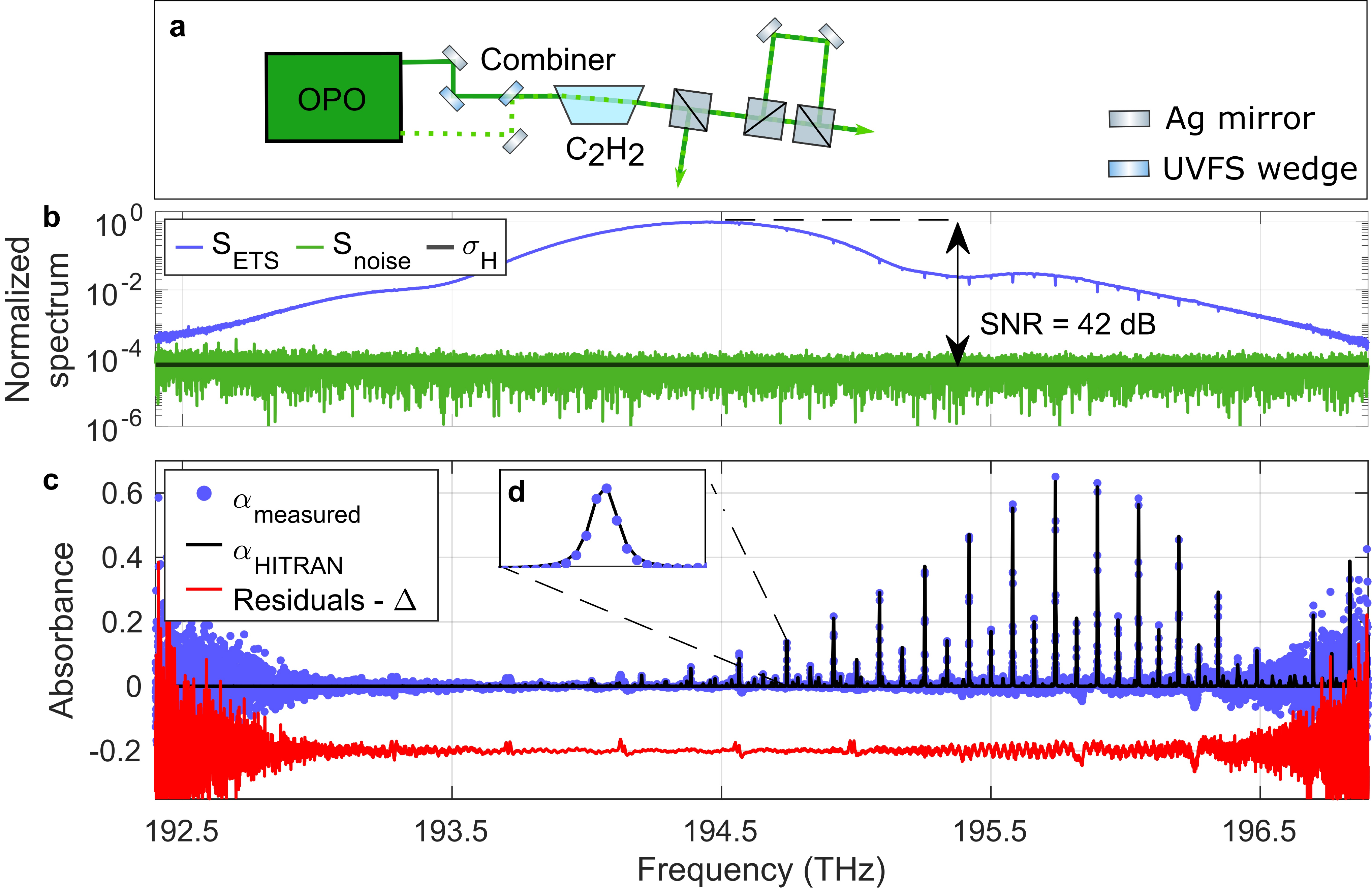}
\caption{
Coherently-averaged 2.15-second-long DCS measurement of a low-pressure acetylene (\acetyleneFormula{}) sample cell (Thorlabs, GCA20) with nominal values $L_\mathrm{nom}=14 \pm 3$~\unit{\milli\meter} and $p_\mathrm{nom}=66 \pm 13$~\unit{\milli\bar}. \\
\subfig{a} Experimental set-up of the DCS measurement. The components are listed in the Methods. 
\subfig{b} Measured transmission spectrum $\etsSpectrum{ETS}$ (blue), and the corresponding noise floor $\etsSpectrum{noise}$ (green). \\
\subfig{c} Comparison of the predicted absorbance based on HITRAN data $\absCoeff{HITRAN}$ (black) and the measured absorbance $\absCoeff{measured}$ (blue). $T=22.8$~\unit{\celsius}, $L=13.4$~\unit{\milli\meter}, $p=67.9$~\unit{\milli\bar}. Residuals (red), plotted with a vertical offset for better visibility, are calculated by subtracting $\absCoeff{HITRAN}$ from $\absCoeff{measured}$. \\
\subfig{d} Zoom-in to a single absorption line of acetylene to illustrate the measurement resolution of 250~\unit{\mega\hertz}. Each marker corresponds to an optical frequency comb line. 
}
\label{fig:DCS_signal}
\end{figure}

In this section, we present a SWIR DCS measurement with the OPO signal combs. The measurements are taken at the nominal OPO configuration as described above. As illustrated in Fig.~\ref{fig:DCS_signal}a, the two signal combs are combined on an uncoated ultra-violet fused silica (UVFS) wedged window. The combined beams are sent through a 14-\unit{\milli\meter}-long sample cell filled with acetylene (\acetyleneFormula{}) at a nominal pressure of 65~\unit{\milli\bar}.  

\begin{figure}[H] % [h]%
%\centering
\includegraphics[width=0.4\textwidth]{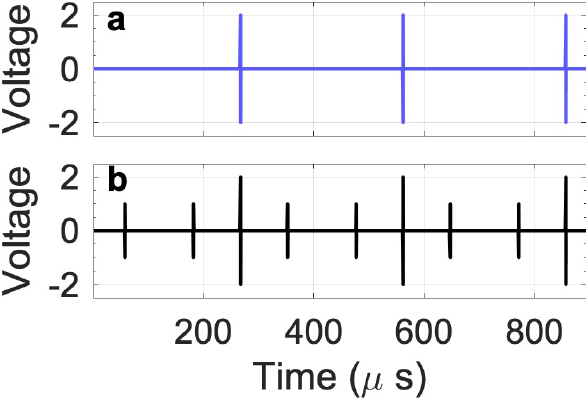}
\caption{
Example signal interferograms. \subfig{a} DCS interferogram used for spectroscopy analysis. \subfig{b} Reference signal containing three interferograms per \Dfrep{} period, used to effectively triple the phase sampling rate.
}
\label{fig:signal_stacking}
\end{figure}

The data structure in each channel corresponds to interferograms (IGMs): in time intervals defined by $1/\Delta f_{\mathrm{rep}}\approx 300~\unit{\micro\second}$, the pulses from the two combs become temporally overlapped and an interferometric signal is generated. To help with the idler measurements that will be discussed in the next subsection, we wanted to operate with \Dfrep{} as small as possible while still supporting coherent averaging. The pump laser supports coherent averaging at $\Delta f_{\mathrm{rep}}\approx 3~\unit{\kilo\hertz}$ due to its robust prototype housing, and the OPO inherits its low timing jitter~\cite{Dudley1994, Bauer2022}. However, since the OPO is constructed on a breadboard setup it is more sensitive to noise and would need $\Delta f_{\mathrm{rep}}\approx 10~\unit{\kilo\hertz}$. Therefore, to operate at a \Dfrep{}=\DfrepNominal{} while still sampling the OPO phase fluctuations fast enough, we implemented the simple pulse stacking setup shown in Fig.~\ref{fig:DCS_signal}a. A small portion of the signal beams are split 50/50 on a beam splitter, and one port is delayed by about a third of the round-trip cavity length before being recombined with the other port, resulting in three IGM patterns within each \Dfrep{} period. Details of the data acquisition and the processing steps to make use of these additional IGMs are explained in the Methods (see Fig. \ref{fig:signal_stacking}). The data from both channels (one with and one without the pulse stacking) are recorded on a data acquisition card.

In the coherent averaging process, each 1/\Dfrep{} period $k$ of the data is frequency-shifted and interpolated onto an optical delay grid of duration 1/\frep{}. This yields a complex array $\etsTime{k}(t)$ for a data period $k$. The coherently averaged signal is given by 
\begin{align}
\label{eq:y_ets}
    \etsTime{ETS}(t)=\frac{1}{N}\sum_{k=1}^{N}\etsTime{k}(t).
\end{align}
where ETS stands for equivalent time sampling. To obtain the noise floor we use the same set of arrays, but summed with alternating sign so that coherent signals cancel and only the noise remains:
\begin{align}
\label{eq:y_alternating}
    \etsTime{noise}(t)=\frac{1}{N}\sum_{k=1}^{N}(-1)^k \etsTime{k}(t),
\end{align}
where $N$ is implicitly even in Eq.~\ref{eq:y_alternating}. The DCS spectrum is given by $\etsSpectrum{ETS}(\nu)=|\mathscr{F}\left[\etsTime{ETS}\right](\nu)|$ and the noise spectrum is given by $\etsSpectrum{noise}(\nu)=|\mathscr{F}\left[\etsTime{noise}\right](\nu)|$. This approach to calculating the noise floor is a generalization of Fourier transforming a long time trace consisting of multiple periods and looking mid-way between the RF comb lines\cite{Yan2017, Nitzsche2021}. 

Figure~\ref{fig:DCS_signal}b shows the coherently averaged results from a 2.15-\unit{\second} duration measurement. The nominal noise floor is determined as the mean of $\etsSpectrum{noise}$ for frequencies within the FWHM bandwidth of $\etsSpectrum{ETS}$. We thereby obtain a peak sensitivity of $\sigma_H=5.9\times 10^{-5}$, which corresponds to a \SNRsqrtT{} of \signalSNRnorm{}~\dBRootHz{}. The transmission dips due to absorption in the acetylene cell are also clearly imprinted on the DCS spectrum. The full range of frequencies shown in the figure, where the spectrum is above its noise floor, is $\approx 4.5~\unit{\tera\hertz}$. Given the repetition rate difference \Dfrep{}=\DfrepNominal{}, this translates to about 60~\unit{\mega\hertz} RF bandwidth (see Eq.~\ref{eq:RFbandwidth}), which is significantly below \frep{}/2. Therefore, it was straightforward to position the RF spectrum in a way to avoid spectral aliasing.

% section~\ref{methods:absorbance}

In Fig.~\ref{fig:DCS_signal}c the transmission features are converted to absorbance $\absCoeff{measured}$ using the approach to infer $\absCoeff{}$ from $\etsTime{ETS}(t)$ described in the Methods. A comparison between measured and predicted absorption $\absCoeff{HITRAN}$ is also shown. The prediction is based on the HITRAN database~\cite{Kochanov2016,Gordon2022} and accounts for the cell's pressure $p$, temperature $T$, and an off-center beam path $L$ through the sample cell that falls within the provided tolerances. As our measurements are conducted without an optical reference or exact knowledge of $\Delta f_{\mathrm{CEO}}$, we cannot independently link RF frequencies to optical frequencies. Therefore, we rely on the unique pattern of the absorption features to align the measured and the predicted absorbance by a center frequency shift.   

We obtain generally good agreement in the shape and strength of the absorption features, as indicated by the residuals. There is a periodic disturbance in the residuals that is not due to absorption. This effect is likely due to intra-cavity beam height changes in the OPO cavity which lead to small polarization rotations on the combs, since beam rotations can turn into etalon-like delayed electric field components upon propagation through the birefringent PPLN crystal. We used a relatively high OPO output coupling rate (12\%) to mitigate intracavity enhancement of these and any other etalon effects. The residuals have a standard deviation of $7.7\times10^{-3}$ for the spectral range 193.2~\unit{\tera\hertz}$^{-1}$ to 195.8~\unit{\tera\hertz}$^{-1}$. This standard deviation becomes  $4.6\times10^{-3}$ when the depolarization peaks in the above range are excluded.

We calculate the FOM of this measurement according to Eq.~(5) in~\cite{Guay2023} and obtain a competitive FOM~=~$3.5\times 10^7$~\unit{\hertz}$^{1/2}$. For the given average power level and amplification the shot-noise is estimated to be 1440~\nvsqrtHz{}. With 3000 comb lines ($\Delta\nu_{\mathrm{FWHM}}/f_{\mathrm{rep}}$), a detector noise-equivalent power (NEP) of $2.0\times10^{-14}$~\unit{\watt}/\unit{Hz}$^{1/2}$, RIN below -160~\dBcHz{} at the relevant \unit{\mega\hertz} offset frequencies, and noise contributions from the digitization card of 1380~\nvsqrtHz{} (inferred from the datasheet), a theoretical sensitivity $\sigma_H$ of $1\times 10^{-4}$~\unit{\second}$^{1/2}$ (\SNRsqrtT{}=$1/\sigma_H=40.0$~\dBRootHz{}) can be expected, where the noise contribution from the digitization unit and shot-noise are of similar magnitude. The calculated \SNRsqrtT{} is in good agreement with the measured value.

\subsection*{Intracavity cross-comb spectroscopy setup}
\label{sec:results_CCS_setup}

Next, we discuss the mid-IR spectroscopy measurements and show how the high average powers provided by the OPO can be directly translated into enhanced sensitivity. To allow for a compact description of the new approach described in this subsection, we denote the two pump combs as \pump{1} and \pump{2}, the signal combs as \signal{1} and \signal{2}, and the idler combs as \idler{1} and \idler{2}. 

\begin{figure}[H] % [h]%
\centering
\includegraphics[width=0.9\textwidth]{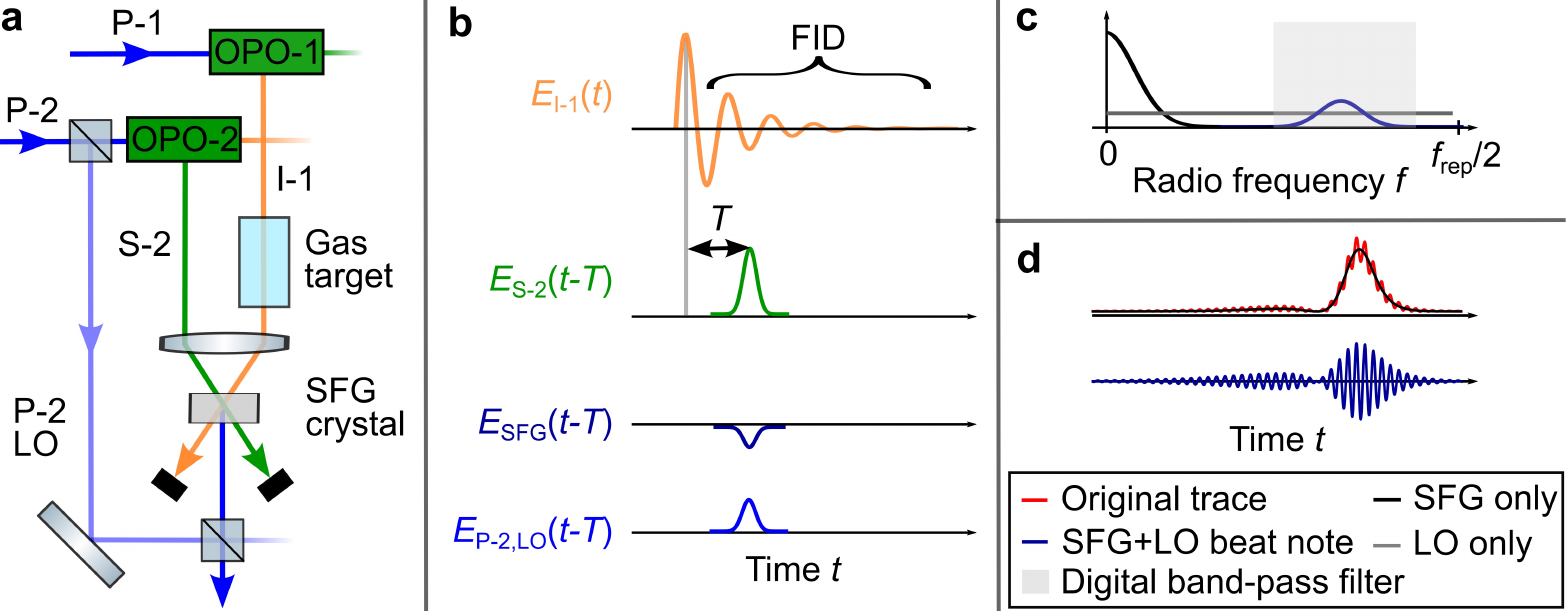}
\caption{
Concept of the OPO-based CCS approach. \\
\subfig{a} Minimal implementation of CCS on the basis of two OPOs. \\
\subfig{b} Temporal profiles of the interacting waves. From top to bottom: the FID of the idler trace; the signal pulse at a delay $T$ probing the corresponding portion of the FID; the SFG wave from the product of the idler and signal wave generated at delay $T$; the temporally overlapped \pump{2} LO. \\
\subfig{c} Components of the RF spectrum: the SFG intensity (black), the LO shot noise (grey), digital band-pass filter (light grey), and the product between the LO and SFG fields (dark blue; heterodyne signal). \\
\subfig{d} Conceptual illustration of the time-domain oscilloscope traces (compare equivalences in c). The absorption features have negative delays with respect to the center burst because we assume the idler FID is sampled by the ``slower" signal comb. Red: original trace from the SFG+LO mixing; Black: low-pass filtered trace containing only the SFG part; Dark blue: band-pass filtered trace containing only the SFG+LO heterodyne signal, which contains the spectroscopic information. 
}
\label{fig:CCS_concept}
\end{figure}

As mid-IR DCS measurements are typically limited by the low saturation power and high noise of Peltier-cooled mid-IR detectors, direct mid-IR DCS does not take full advantage of the low-noise and high-power idler beams the OPO provides. To overcome this limitation we use CCS, which is a heterodyne up-conversion technique first demonstrated in~\cite{Liu2023}. To help explain our new approach to this technique, we first illustrate in Fig.~\ref{fig:CCS_concept}a a minimal implementation of CCS that is enabled by our tunable dual comb OPO. The temporal profiles of the interacting optical waves are illustrated in Fig.~\ref{fig:CCS_concept}b. The detection process involves the following main steps: 

\begin{enumerate}
    \item \idler{1} is sent through an absorbing gas, leading to temporally-delayed features on the electric field (the so-called free induction decay, FID. See top of Fig.~\ref{fig:CCS_concept}b).
    \item \idler{1} is mixed with \signal{2} in a sum frequency generation (SFG) crystal, thereby sampling the FID of \idler{1}.
    \item The generated SFG wave $E_\mathrm{\mathrm{SFG}}$ has the same optical carrier frequency as the pump. Therefore, for subsequent heterodyne detection, it is combined on a beam splitter with a small portion of \pump{2} that acts as the local oscillator (LO) wave. 
    \item The combined beam (SFG and LO) is detected on an InGaAs photodiode. This results in three terms: the intensity of the LO wave, the intensity of the SFG wave, and the heterodyne beat signal corresponding to the product between the SFG and LO electric fields. The heterodyne signal is the one of interest as it contains the spectroscopic information. 
\end{enumerate}

% section \ref{methods:CCS_theory}

A theoretical description of the CCS signal is provided in the Methods. The RF spectrum of the photodiode signal is illustrated in Fig.~\ref{fig:CCS_concept}c, and the corresponding time-domain features are shown in Fig.~\ref{fig:CCS_concept}d. The black lines correspond to the intensity of the SFG wave, which has a strong time-domain peak near $T=0$ and decays quickly away from $T=0$. The grey line indicates the LO including its white shot noise contribution. The blue lines show the heterodyne signal of interest, which is analogous to a dual-comb interferometry signal. With our repetition rate difference of \DfrepNominal{}, the heterodyne signal and SFG signal are narrow-band enough to occupy separate non-overlapping regions of the RF space without spectral aliasing issues, and can therefore be easily extracted using either analog or digital band-pass filters.

\begin{figure}[H] % [h]%
\centering
\includegraphics[width=0.9\textwidth]{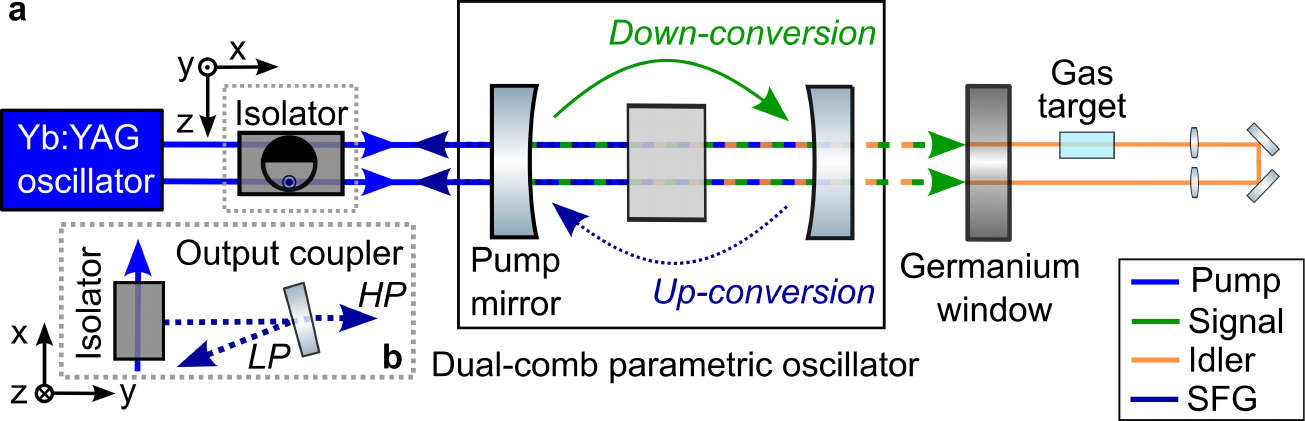}
\caption{
\subfig{a} Side view of the experimental implementation of the CCS scheme. Note that we only show a simplified illustration of the OPO cavity (compare Fig.~\ref{fig:OPO_schematic}). After the dual-comb OPO cavity, a Germanium window blocks any signal or pump beam components co-propagating with the output idler beams. Each idler beam passes through the gas target either on its path from or towards the OPO. Lenses are used to create a focus of each idler beam at a common position serving as a point of symmetry where the idler propagation is reversed. Each idler propagates on the path of the other idler beam to the intra-cavity PPLN crystal where the idler and signal combs are mixed in an SFG process. At the pump mirror, the SFG wave (dashed, dark blue) is interfered with a small residual pump reflection serving as LO. The rejection port of a 5-\unit{\milli\meter}-aperture isolator (InPut Optica, HPTG) is used to separate the backwards propagating CCS signal from the incoming pump. One of the two comb paths is blocked, and the other SFG+LO field is used for detection. \\
\subfig{b} Top-view of the detection set-up where the CCS wave is split by a partially transmitting optic into a low-power (LP) and high-power (HP) channel.
}
\label{fig:CCS_implementation}
\end{figure}

Our dual-comb OPO enables a nearly ideal architecture for CCS, which is depicted in Fig.~\ref{fig:CCS_implementation}a-b. The key idea is that \idler{1} is sent through an absorbing target (step 1) and then simply routed back along the path of \idler{2}. In this configuration, \idler{1} passes back through the intra-cavity PPLN crystal at the same position as \idler{2} but counter-propagating. As the collinearly-pumped OPO cavity is linear, \signal{2} is also counter-propagating intra-cavity at the same position. Hence, SFG occurs between \idler{1} and \signal{2} (step 2). This SFG process is inherently efficient, phase-matched, aligned, and spectrally overlapped with \pump{2} since it simply represents the reverse of the optical parametric amplification process. Furthermore, since SFG corresponds to the time-domain product between the electric fields of \idler{1} and \signal{2}, the generated SFG wave is always synchronized with the intra-cavity \signal{2} pulses. Given that the timing of \pump{2} and \signal{2} is locked together by the OPO process, the SFG wave is also temporally overlapped with the incoming \pump{2} pulses. Thus, a small residual reflection of \pump{2} by the HT$_{pump}$ / HR$_{signal}$-coated surface of the input mirror ($R_{pump}<1$\%) inherently functions as the \pump{2} LO beam (step 3). Thus, no additional components or alignment is needed to obtain interference between the SFG wave and the \pump{2} LO. In short, all six important constraints to obtain the heterodyne beating are inherently satisfied (SFG phase-matching and efficiency; LO generation; spatial, spectral and temporal overlap with the LO). Since the (SFG+LO) beam of interest is counter-propagating to the incoming pump light, an isolator is added to the setup to extract the SFG+LO beam via the isolator’s rejection port. Note that a second beam is present with the roles of the combs interchanged (\ie{} \pump{1} acting as the LO instead of \pump{2}). This beam is blocked in our setup, but it could potentially be used to improve the SNR by a factor of 2.

\subsection*{CCS measurements}
\label{sec:results_CCS_measurements}

To show the ultra-high sensitivity measurements supported by combining our dual-comb OPO with intracavity CCS detection, in this section we present a measurement on ambient air with the idler tuned to 3320~\unit{\nano\meter} to probe strong methane absorption features. The light reflected by the isolator is split into two free space paths: a high-power (HP) path containing $>$95\% of the power, and a low-power (LP) path that is attenuated to about 0.5\%. The paths are illustrated conceptually in Fig.~\ref{fig:CCS_implementation}b. Both paths are coupled into separate single-mode fibers to implement a customized high-sensitivity detection scheme. In the HP path, we measure a fiber-coupled LO average power of 4.65~\unit{\milli\watt} and, when the \idler{1} and \signal{2} pulses are synchronous at a delay $T=0$, we estimate a fiber-coupled SFG power of 55~\unit{\milli\watt} (0.22~\unit{\nano\joule}).

\begin{figure}[H] % [h]%
\centering
\includegraphics[width=0.9\textwidth]{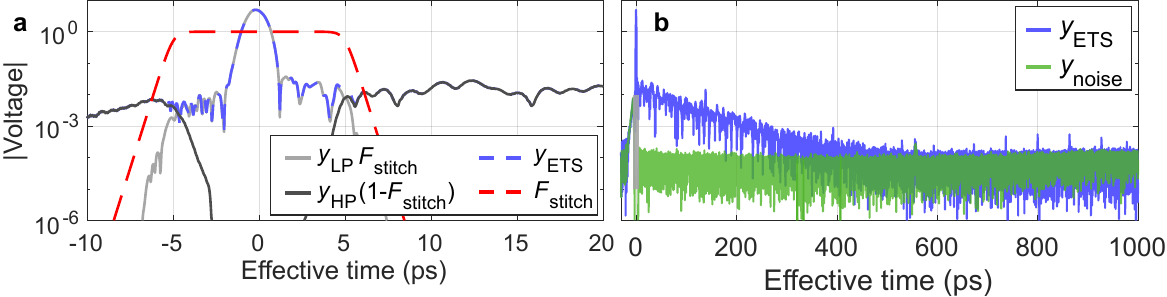}
\caption{
Coherently averaged CCS signals from the LP and HP channels.\\
\subfig{a} Magnitude of the LP, HP, and composite traces zoomed in to the center burst. The peak of the LP channel is scaled such that its FID matches that of the HP channel. The composite trace is derived by summing the product of the $\etsTime{ETS}(t)=\etsTime{LP}(t)F_\mathrm{stitch}(t)+\etsTime{HP}(t)(1-F_\mathrm{stitch}(t))$. \\
\subfig{b} Magnitude of $\etsTime{ETS}$ (blue) and its noise floor $\etsTime{noise}$ away from $T=0$ (green). 
}
\label{fig:CCS_stitching}
\end{figure}

The strong yet low noise signals in our setup present a unique measurement challenge: a high photocurrent of order 100~\unit{\milli\ampere} must be detected while supporting a high bandwidth $>$100~\unit{\mega\hertz} and a measurement noise floor of around 100~\unit{\nano\ampere} (to resolve shot noise of the 4-\unit{\milli\watt} LO beam). Most photodiodes cannot handle such a large current without saturation, and high-speed data acquisition systems generally do not have a dynamic range of $10^6$. To overcome these issues we combine data from the LP and HP paths as follows. The LP path is attenuated in order to accurately measure the center burst without saturation. The HP path is configured so that the $T=0$ center burst is strongly saturated, but with enough sensitivity to reach shot-noise limited performance away from $T=0$. To avoid degrading the FID signal due to time-delayed saturation effects on the HP channel, we configure the system so that the raw time trace is ``backwards” (see Fig.~\ref{fig:CCS_concept}d), as this means that saturation-induced distortions occur at negative optical delays and can be discarded. The two channels are coherently averaged individually and then combined digitally to obtain the full signal (composite trace). Details of the stitching process and the detection setup are in the Methods. The coherently averaged LP and HP signals are shown in Fig.~\ref{fig:CCS_stitching}a. The same terminology from Eqs.~\ref{eq:y_ets} and \ref{eq:y_alternating} are used for the signal and noise floor, respectively. The amplitude and phase of the LP channel have been carefully scaled to match the HP channel within the FID region, which permits digital stitching of the two signals together. The stitched signal and its noise floor are shown in Fig.~\ref{fig:CCS_stitching}b, which illustrates the decay of the FID towards the signal's noise floor. 

 % section~\ref{methods:CCS_detection}
 
\begin{figure}[H] % [h]%
\centering
\includegraphics[width=0.9\textwidth]{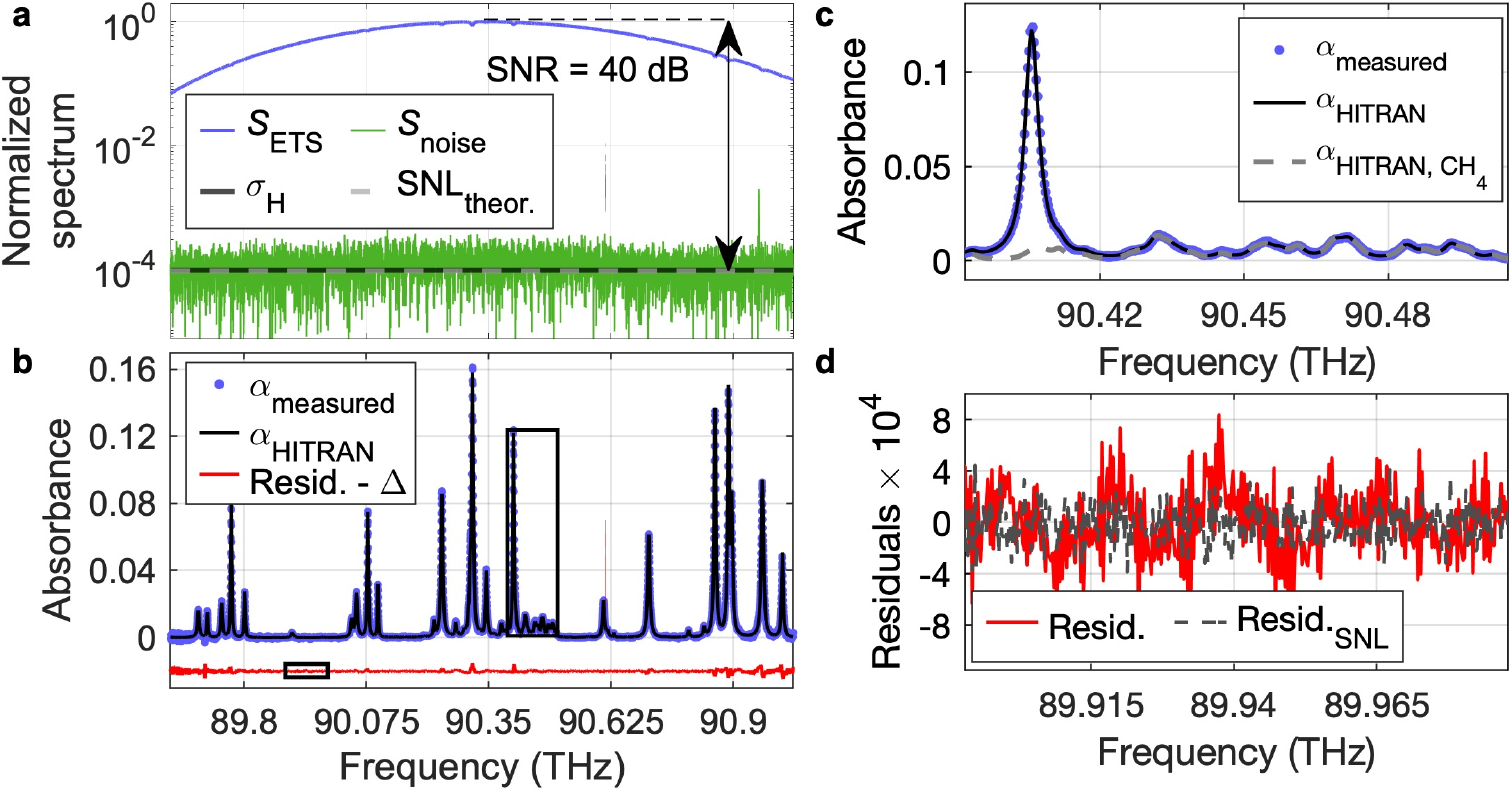}
\caption{
Coherently-averaged 10-\unit{\milli\second}-long CCS measurement of the ambient laboratory air. \\
\subfig{a} Measured spectrum (blue) and its noise floor (green). The blue curve is obtained from $\etsTime{ETS}(t)$ over the time range from -20~\unit{\pico\second} to 2800~\unit{\pico\second}. The green curve is obtained from $\etsTime{noise}(t)$ over the full time range except for the small region around $T=0$ where the stitching occurs at 5~\unit{\pico\second}. Identical temporal filtering has been applied to $\etsTime{ETS}(t)$ and $\etsTime{noise}(t)$ to ensure that the latter curve represents the noise floor of the former curve. Averaged noise floor (black). Theoretical noise floor (grey) calculated from shot noise of the LO, with a correction to account for the reduced detector responsivity at the center RF frequency of the CCS interferogram.  \\
\subfig{b} Measured absorption spectrum (blue) compared with HITRAN calculations (black). The residuals are shown by the red, which is displaced from 0 for visibility.  \\
\subfig{c} Zoom-in on the absorption near the methane features to highlight the trace-gas detection aspects of the measurement. $\absCoeff{HITRAN}$ is calculated with a sampling rate corresponding to the resolution of the measured data. \\
\subfig{d} Zoom-in on the residuals where there are no water or methane features, for comparison to the noise floor.
}
\label{fig:CCS_results}
\end{figure}

Figure~\ref{fig:CCS_results} shows the spectroscopic features of the measurement. The spectrum has an \SNRabsolute{} of 40.2~\unit{\decibel} ($\sigma_H=9.7\times 10^{-5}$) as shown in Fig.~\ref{fig:CCS_results}a. Given the short measurement time of 10-\unit{\milli\second}, this corresponds to a high \SNRsqrtT{} of \bestSNR{}~\dBRootHz{}. Accounting for the spectral shape of the signal \cite{Guay2023}, we obtain an ultra-high FOM of \bestFOM{}~\unit{\hertz}$^{1/2}$. The shot-noise limited noise floor is shown by the grey dashed line, indicating that the measurement is close to its fundamental sensitivity limit. 

The absorption spectrum is determined as described in the Methods and is shown in Fig.~\ref{fig:CCS_results}b. The measured spectrum is in good agreement with the HITRAN database, as indicated by the small residuals. The standard deviation of the residuals is $5.6\times 10^{-4}$ in a 1.2~\unit{\tera\hertz} wide segment centered about the peak. To obtain the best overlap, we account for the temperature $T$, pressure $p$, as well as the length $L$ of the idler beam path and the relative abundances $F$ of the detected molecules: $T=25$~\unit{\celsius}, $p=0.95$~\unit{\bar}, $L=271$~\unit{\centi\meter}, $F_{\mathrm{H_2O}}=0.01$, $F_{\mathrm{CH_4}}=2.5$~ppm. Figure~\ref{fig:CCS_results}c shows an inset of the absorption spectrum in the region of the methane absorption band, highlighting the trace-gas detection of methane at the 1~ppm level. The figure shows a coherently averaged measurement over 10~\unit{\milli\second}, but the achieved sensitivity is high enough such that the absorption features can already be clearly resolved in a single IGM period of 300~\unit{\micro\second}. Figure~\ref{fig:CCS_results}d shows an inset of the residuals in a spectral region where there is negligible absorption from water vapor or methane. The directly calculated residuals are close to the noise floor-limited residuals. This shows that even the absorbance, despite its sensitivity to small distortions in the coherent signal from unsuppressed etalons, has close to shot-noise limited sensitivity.

In addition to the above measurement of ambient air, which was chosen to highlight the achievable sensitivity and trace-gas detection of methane, two further measurements are presented in the Supplementary information. There, we present measurements on a low-pressure acetylene cell at two different wavelength tuning points (2960~\unit{\nano\meter} and 3790~\unit{\nano\meter}) to show the high-resolution and wavelength-tunability capabilities of the system.

\section*{Discussion}

We demonstrate a novel platform for high-resolution spectroscopy in the SWIR and mid-IR spectral regions that combines a simple system architecture with high sensitivity. Our results are based on a new light source consisting of a tunable dual-comb OPO pumped by a picosecond dual-comb solid-state laser. The laser and the OPO cavities use spatial multiplexing to generate both combs in a single cavity. This means all the optical elements are shared both in the laser cavity and in the OPO cavity, leading to high mutual coherence between the two combs from a simple setup. The implementation of the spatial multiplexing in the OPO cavity allows to synchronously pump both OPO combs simultaneously at a repetition rate difference of several ~\unit{\kilo\hertz} while maintaining high spectral overlap between the two combs. Hence, phase tracking of the heterodyne dual-comb signals and subsequent coherent averaging is possible even in fully-free running operation of both cavities.

While maintaining a narrow instantaneous bandwidth, the OPO covers a broad spectral range due to its tuning capability. The OPO signal tuning range is from  \SignalTuningMin{}~\unit{\nano\meter} to \SignalTuningMax{}~\unit{\nano\meter}, and the OPO idler tunes from \IdlerTuningMin{}~\unit{\nano\meter} to \IdlerTuningMax{}~\unit{\nano\meter}. This wide spectral range in the SWIR and mid-IR allows for the investigation of absorption features from numerous molecules. The high output power in combination with the narrow spectral bandwidth results in high power per comb line of up to 210~\unit{\micro\watt} in the SWIR and 160~\unit{\micro\watt} in the mid-IR. Furthermore, the dual-comb system has low relative-intensity noise measured to be below -160~\dBcHz{} at megahertz offset frequencies. In the context of direct DCS measurements, the high power of the OPO makes it particularly compelling for measurements involving high losses such as long open paths or in broadband multi-pass cells. 

We carried out two spectroscopic measurement campaigns. In the SWIR measurements, we employ the signal to show the compatibility of the OPO with high-resolution and high-sensitivity direct DCS measurements. In the other campaign, we focus on the idler to show the fundamental boost in sensitivity achievable by operating the OPO as a spectroscopic transceiver while using the full average power available.

The shot-noise limited SWIR measurements on a low-pressure acetylene cell had a \SNRsqrtT{} of \signalSNRnorm{}~\dBRootHz{} and a FOM of $3.5\times10^7$~\unit{\hertz}$^{1/2}$. For comparison, a \SNRsqrtT{} of 41~\dBRootHz{} was obtained with electro-optic dual-combs~\cite{Millot2016}, but that system had fewer comb lines (115) and a limited tuning range of 100~\unit{\nano\meter}. In another experiment, a FOM of $7.2\times10^7$~\unit{\hertz}$^{1/2}$ was obtained by sending 23.5~\unit{\milli\watt} to a photodetector that is fast enough to allow for correction of the introduced non-linearity~\cite{Guay2023}. The measurement in \cite{Guay2023} was limited by the RIN of the used laser source, so there is room for further improvement by combining high-power detectors with bulk solid-state laser-based light sources that are not limited by RIN.

In order to utilize the high idler powers provided by the OPO and overcome the limitations posed by mid-IR detectors, we demonstrated a new cross-comb spectroscopy scheme based on up-conversion between the idler and the intra-cavity signal. The approach, which requires only a few extra parts compared to the source itself, yielded a record-high FOM of \bestFOM{}~\unit{\hertz}$^{1/2}$ and trace-gas detection of methane in ambient air. We showed coherently averaged results, but the sensitivity is high enough to already resolve methane absorption in a single interferogram period. Our results represent the first implementation of OPO-based CCS and the first intracavity CCS architecture. Here, the OPO is operated as a transceiver, such that the intracavity signal automatically up-converts the idler free induction decay (FID) to intrinsically be spectrally and temporally overlapped with the pump beam which then can serve as the local oscillator. To the best of our knowledge, we demonstrated the highest FOM for high-resolution mid-IR DCS without relying on cavity enhancement~\cite{Muraviev2018, Ycas2018, Chen2020, Komagata2023,Vasilyev2023, Liu2023, Konnov2023}. The sensitivity far exceeds the fundamental shot-noise limited performance that would be possible with direct DCS using the same 4-\unit{\milli\watt} local oscillator power. The improvement comes from the time-gating effect of the up-conversion, where away from $T=0$, the main idler pulses do not get up-converted and therefore they do not contribute to shot noise~\cite{Winzer1997}; the dominant contribution to shot noise is instead the lower power \pump{2} LO beam. In fact, reaching similar performance in a direct DCS configuration would require using the full mid-IR power (300~\unit{\milli\watt}) on a photodiode with near-ideal quantum efficiency. This exceeds the capabilities of mid-IR detectors by two to three orders of magnitude. One other important point is that, while measurements with the OPO signal exhibited small spectral modulations attributable to intracavity depolarization effects, these effects are not visible on our measurements with the OPO idler since any time-delayed components on the signal do not get efficiently down- or up-converted. 

To take full advantage of the scheme we needed to split the measurements into separate low- and high-power channels. This two-channel detection may be necessary in general when targeting shot-noise-limited performance due to the very stringent demands placed on the photodiode and electronics in terms of power, dynamic range, and bandwidth. In our measurements, the two channels can be stitched together with small residuals because we carefully calibrated the electronic transfer functions of both channels. Also, the demands on the matching of the two channels are reduced when the goal is to detect very weak features resulting in a weak FID, which is our primary motivation. As soon as high sensitivity is not critical, e.g. when sampling strong absorption features, it is not necessary to push the measurement to its fundamental shot noise limits. In that case, a single detector operated at moderate power would be sufficient. For measurements using the full power of the OPO, further sensitivity improvements could be obtained by OPO power scaling. Eventually, the damage threshold of the photodiode or the electronic amplifier would set a limit to the useable power, but there is no indication that we are close to such a limit with the current measurements. 

In conclusion, our results validate the use of tunable OPOs as an attractive way forward for dual-comb spectroscopy applications. Future work will involve extending to longer wavelengths and deploying the set-up for the detection of trace-gas mixtures in multi-pass cells or longer air paths. The obtained SNR would support combline-resolved measurements of absorbance at the $<1~$ppm level after coherent averaging timescales of one minute. For measurements in ambient conditions, higher repetition rates of around 1~\unit{\giga\hertz} could be used to obtain even higher SNR. For high-resolution applications, the repetition rate could be reduced, or spectral interleaving techniques could be deployed. The potential of low-noise dual-comb solid-state lasers and OPOs to reach repetition rates from below 100~\unit{\mega\hertz} to $>1$~\unit{\giga\hertz} while maintaining ultra-low noise performance make them ideally suited to access this parameter space.

%TC:ignore
\section*{Methods}

\subsection*{Theoretical description of CCS signal}
\label{methods:CCS_theory}

In the following, we use the same notation as introduced in the experimental results section. To extract an absorption spectrum from a CCS measurement, it is useful to express the electronic signal in terms of the properties of the combs and the sample. For this, we determine the charge generated per pulse on the photodiode as a function of delay between the two mixing combs. The SFG electric field $E_\mathrm{SFG}$ can be approximated as:

\begin{align}
    E_\mathrm{SFG}^{(n)}(t) = e^{in\delta_{is}} E_\mathrm{\idler{1}}(t+T_n)E_\mathrm{\signal{2}}(t),
\end{align}

where the idler field $E_\mathrm{\idler{1}}$ implicitly contains the spectroscopic information, $n$ represents a pulse index, $\delta_C=\phi_\mathrm{\idler{1}} + \phi_\mathrm{\signal{2}} - \phi_\mathrm{\pump{2}}$ is a carrier envelope phase (CEP) shift term with $\phi_\mathrm{j-k}$ being the CEP shift of wave $j$ of comb $k$ from one pulse to the next, and $T_n$ is the delay between the combs at pulse pair $n$. The envelopes $E_j$ are the positive-frequency electric fields of each wave with the CEP set to zero. To obtain a photodiode signal, we interfere this SFG beam with a (small) replica of pump \pump{2} (LO), which results in a time-varying photocurrent source varying on ultrafast timescales:
\begin{align}
    i_\mathrm{\mathrm{source}}^{(n)} = \eta\left[
    \left| E_\mathrm{SFG}^{(n)}(t)\right|^2
    +
    \left| E_\mathrm{\pump{2}}(t)\right|^2
    +
    2\mathrm{Re}\left(e^{in\delta_C} E_\mathrm{SFG}^{(n)}(t)E_\mathrm{\pump{2}}^*(t)\right)
    \right]
\end{align}
where $\eta$ is the photodiode responsivity. Since the photodiode has a ``slow” response, we end up integrating this photocurrent source to obtain a charge per pulse-pair:
\begin{align}
    Q_n = Q_\mathrm{SFG}(T_n) + Q_\mathrm{pump} + 2\mathrm{Re}\left(Q_\mathrm{CCS,n}\right).
\end{align}
Here, $Q_\mathrm{SFG}$ corresponds to a convolution between the envelopes of the idler and signal fields; it is strong when the main pulses of each wave are temporally overlapped ($T_n=0$). $Q_\mathrm{pump}$ is a constant term corresponding to the pump pulse energy at the photodiode. The CCS term $Q_\mathrm{CCS}$ is phase sensitive and given by:
\begin{align}
    Q_\mathrm{CCS,n}=e^{in\delta_C}\int E_\mathrm{\idler{1}}(t+T_n)E_\mathrm{\signal{2}}(t)E_\mathrm{\pump{2}}^*(t)dt,
\end{align}
where the integration limits of infinity implicitly assume that neighboring pulses in the train are neglected. To elucidate further, note that the signal-pump product within the integrand constitutes a gate with which the idler field is convolved. By approximating the phase and delay variables as continuous (a valid assumption provided the repetition rate difference is small enough to avoid aliasing) and multiplying by the repetition rate to obtain a current, we obtain:
\begin{align}
    i_\mathrm{CCS}(T)\approx f_{\mathrm{rep}} \exp(2\pi i \nu_{CCS} T)\int E_\mathrm{\idler{1}}(t+T) E_\mathrm{\signal{2}}(t)E_\mathrm{\pump{2}}^*(t) dt,
\end{align}
where the effective carrier frequency of this function on optical timescales is $\nu_{CCS}=(f_{\mathrm{rep}}/\Delta f_{\mathrm{rep}})(f_{\mathrm{rep}}\delta_C/(2\pi))$. 

Performing a Fourier transform, we obtain:
\begin{align}
\label{eq:ccs_main}
    \tilde{i}_{CCS}(\nu) 
    &= \tilde{E}_{\idler{1}}(\nu-\nu_{CCS})\tilde{G}(\nu - \nu_{CCS})^* 
    \nonumber \\
    &= \tilde{E}_{\idler{1}, init}(\nu - \nu_{CCS}) \tilde{H}(\nu - \nu_{CCS}) \tilde{G}(\nu - \nu_{CCS})^*
\end{align}
where tilde denotes a frequency domain quantity, $\tilde{E}_{\idler{1},init}$ is the idler electric field before passing the absorptive sample, $\tilde{H}$ is the absorption transfer function, $\tilde{G}(\nu)=\mathscr{F}\left[G(t)\right]$, and the gate $G(t)$ is given by:
\begin{align}
    G(t)=E_\mathrm{\signal{2}}(-t)^*E_\mathrm{\pump{2}}(-t).
\end{align}
Equation~\ref{eq:ccs_main} shows that the CCS spectrum contains the spectroscopic information of the sample, weighted by the spectrum of the idler and the effective gate pulse defined by the product of the signal and pump of the other comb. Therefore, spectroscopic information that lies within the bandwidth of the idler can be retrieved, provided the signal and pump pulses are of comparable duration (which is usually the case for an OPO).

\subsection*{Phase tracking with low \Dfrep{}}
\label{methods:phase_tracking}

Several data processing steps are needed to extract absorption values from the raw data. The overall procedure consists of (i) IGM phase tracking, (ii) coherent averaging, (iii) digital stitching (for CCS measurements only), and (iv) HITRAN fitting, envelope extraction, and etalon removal. Here we discuss step (i); steps (ii)-(iv) are discussed later in the Methods.

%Steps (i) and (iv) are discussed in detail later in the Methods. 
 
We operate at \Dfrep{}=\DfrepNominal{}, which is large enough to coherently average the pump but not the OPO signal \cite{Hebert2019}. The OPO-signal pulse stacking setup of Fig.~\ref{fig:DCS_signal} mimics the role of a larger \Dfrep{} by sampling the phase fluctuations three times more rapidly. 

More specifically, we obtain three arrays of phase samples $\igmPhi{j}$, where $\igmPhi{j}[n]$ is the carrier envelope phase (CEP) of pulse $j\in\{1,2,3\}$ within overall IGM period $n$. The timing difference between these IGMs is stable but not known precisely. Therefore, we use differences between their CEP values to determine the phase fluctuations of the ``main" IGM's $\igmPhi{1}$. To unwrap $\igmPhi{1}$ we calculate 
$\igmDelta{21}=\unwrap{\igmPhi{2}-\igmPhi{1}}$, 
$\igmDelta{32}=\unwrap{\igmPhi{3}-\igmPhi{2}}$, and 
$\igmDelta{13}[n]=\unwrap{\igmPhi{1}[n+1]-\igmPhi{3}[n]}$. $\igmPhi{1}$ is then obtained by taking the sum $(\igmDelta{21}+\igmDelta{32}+\igmDelta{31})$ and removing any overall $2\pi$ phase slope offset. This procedure works for all the traces, so none have to be discarded. In general, the pulse stacking approach could be avoided, or deployed successfully for much lower \frep{} systems, by using mechanically robust cavity assemblies instead of breadboard setups.

For the idler data (\ie{} the CCS measurements) an additional step is required because the idler phase has fluctuations of similar magnitude to those of the signal while also being influenced by the pump phase. To find the idler phase, we calculate $\varphi=\unwrap{\phi_i+\phi_s-\phi_p}$ on each IGM period (which has sufficiently small fluctuations to unwrap without ambiguities), and replace the directly calculated idler CEP values with $\varphi - \phi_s + \phi_p$. This procedure preserves the original value modulo $2\pi$ while removing phase-wrapping ambiguities. Now having an IGM phase array for all channels, coherent averaging is enabled.

\subsection*{Data processing of the interferometric measurements}

Coherent averaging (step (ii)) proceeds as usual by interpolating the phases as obtained in the previous Methods section on the whole time grid, frequency shifting, and interpolating on the optical delay grid. The last step benefits from the fact that the timing grid is inherited from the pump laser, which is in a robust mechanical prototype with low timing jitter. Coherent averaging is applied to all traces. For the CCS signals, the phase array is obtained from the low-power (LP) channel, and that phase array is used to average both LP and high-power (HP) channels. 

The digital stitching step (step (iii)) involves identifying features on the coherently averaged signals that are shared between the LP and HP channels. For the HP channel, the limiting factor is saturation, while for the LP channel, it is the noise floor. Therefore, the first few hundred picoseconds of the free induction decay (FID) are suitable for this purpose as those features are neither saturated in the HP channel nor within the noise floor of the LP channel. Since the LP and HP electronics are not identical, we carried out a measurement of their electronic transfer functions $H_\mathrm{LP}$ and $H_\mathrm{HP}$ with a separate dual comb laser (a prototype version of the one reported in \cite{Phillips2023}); this captures spectral oscillations arising from reflections of the electronic signals. The LP channel is multiplied by $H_\mathrm{HP}/H_\mathrm{LP}$ prior to stitching. The final output of the above procedures is denoted $\etsTime{ETS}(t)$: it is a signal of duration $1/f_{rep}$ whose Fourier series coefficients correspond to comb line pairs.

\subsection*{Extracting the absorbance from DCS and CCS measurements}
\label{methods:absorbance}

In this section, we describe the data processing performed on $\etsTime{ETS}(t)$ to obtain absorbance data (step (iv)). We perform measurements without a reference and therefore need to use $\etsTime{ETS}(t)$ itself to infer the underlying absorption-free envelope that gives rise to the FID. This process is optimized to reduce the residuals between the measured data and the database prediction. The analysis has three interconnected steps involving digital removal of the absorption features, calculating an envelope, and removing ``etalons". Provided all strong absorption features originate from gases whose lines are well tabulated in a database, the FID can be removed accurately and without information loss:
\begin{align}
\label{eq:y_corrected}
    \etsFreq{corrected}(\nu) = \etsFreq{ETS}(\nu)H(\nu + \nu_0;T,P,L)^{-1}
\end{align}
where $\nu$ is the optical frequency variable of the coherently averaged signal and $\nu_0$ a frequency shift parameters used to match the observed absorption features to the predicted ones (i.e. to estimate absolute optical frequencies). Other fit parameters included in the absorption transfer function $H(\nu)=\exp(-\absCoeff{}(\nu)L)$ are temperature $T$, pressure $p$ and sample length $L$. Once the spectroscopic features are well fit, $\etsTime{corrected}(t)$ has weak or negligible FID, and a few-picosecond-long time gate $g(t)$ can be applied to it to obtain an envelope; we use a hyperbolic tangent shaped filter for this purpose:
\begin{align}
    \etsTime{env}(t)=\etsTime{corrected}(t)g(t)
\end{align}
The time gate ensures that no sharp spectroscopic features are left, and the FID removal before applying the gate suppresses background fluctuations arising from convolution of the gate with the absorption features. The (complex) absorption can now be obtained by division:
\begin{align}
    \absCoeff{}(\nu+\nu_0) = -\ln\left(
    \frac{\etsFreq{ETS}(\nu)}{\etsFreq{env}(\nu)}
    \right)
\end{align}
Initially, $\nu_0$ and the environmental parameters of the transfer function $H$ are not known accurately. The calculated absorption $\absCoeff{}(\nu+\nu_0)$ can be used to estimate these parameters as well. Etalons are removed coherently from the input to Eq. \ref{eq:y_corrected} by applying an inverse optical transfer function approach with amplitude, phase, and delay parameters adjusted to suppress the time-domain peaks. Some etalons that are far from the FID region are removed with a time gate.

\subsection*{Detection system used for the DCS and CCS measurements}
\label{methods:CCS_detection}

In the DCS measurements, the combined combs are detected by a fast photodiode (modified Thorlabs, DET10N2). The photodiode is connected to an 80~\unit{\kilo\hertz} bias-tee (DC output terminated with 50~Ohm) followed by a 98-\unit{\mega\hertz} lowpass filter, a 30-dB amplifier (ENA-220T, RFBay), and analyzed with a 14-bit data-acquisition card (Spectrum instrumentation, M4i.4451-x8).

To maximize the SNR of the CCS measurement, we split the CCS signal in a LP and a HP part. The HP trace is configured to measure the FID with high sensitivity, but the center burst is strongly saturated. Conversely, the LP trace is configured to capture the center burst information that is lost in the HP trace (including its carrier-envelope phase for use in coherent averaging). 

For the LP trace, we attenuate the light rejected by the isolator by a factor of about 200 to avoid saturation and detect the fiber-coupled beam with a gigahertz photodiode (DET01CFC). The attenuation is implemented with a wedged laser output coupler ($T<5\%$) followed by an ND filter. The electronic signal generated by the photodiode is first split with an SMA-T piece. One port is terminated with a 50-Ohm resistor, while the other is low-pass filtered (passband frequency 140~\unit{\mega\hertz}), amplified by 30~\unit{\decibel} (ENA-220T, RFBay), and then connected to the oscilloscope (WavePro 404HD, Teledyne Lecroy) by a short BNC cable. The termination resistor directly after the SMA-T is used instead of a bias tee in order to reduce oscillations on the electronic transfer function that arise due to reflections from the low-pass filter.

The HP path uses the full power available (apart from $<5\%$ split off for the LP path). The light is fiber coupled and sent to a higher-power photodiode (modified Thorlabs DET10N2). The DC part of the signal is split off with a bias-T (cut-off frequency 80~\unit{\kilo\hertz}) and terminated with a 50-Ohm resistor. The AC port of the bias-T is connected to a high-pass filter (passband frequency 27.5~\unit{\mega\hertz}) followed by a low-pass filter (passband frequency 98~\unit{\mega\hertz}), and then through a 23-dB amplifier (LNA-580T, RFBay). The resulting center burst far exceeds the saturation level of the amplifier (which is $\approx\pm 3~\unit{\volt}$), but is within the absolute maximum ratings. On the oscilloscope, the vertical zoom level is optimized to reach as high sensitivity as possible while staying below the absolute maximum input voltage limits. Little center burst information is left on the HP channel, but the FID is resolved to shot-noise limited sensitivity. Subsequent stitching of the LP and HP traces yields the underlying signal that could be obtained with an ideal high dynamic range photodiode and electronics chain.

%\section*{Back matter}
\backmatter 

\bmhead{Funding}
This work was supported by a BRIDGE Discovery Project Nr. 40B2-0\_180933 a joint research programme of the Swiss National Science Foundation (SNSF) and Innosuisse – the Swiss Innovation Agency. This project has received funding from the European Research Council (ERC) under the European Union’s Horizon 2020 research and innovation programme (966718).

\bmhead{Acknowledgements} The authors thank Walter Bachmann from the ETH Physics Department Engineering office for the mechanical design of the dual-comb solid-state laser aluminum housing.

\bmhead{Contributions}
C.P. developed the concept of the project. C.B. planned and built the dual-comb OPO cavity. J.P. planned and built the dual-comb solid-state laser. B.W. and J.P. assisted in experimental optimization steps of the laser. B.W. developed the inverted-biprism. M.B. explored suitable experimental configurations of the OPO. C.B. and Z.B. carried out the final OPO alignment and performed the non-spectroscopic experiments. C.P. developed the computer program for coherent averaging and adaptive sampling of the spectroscopy measurements, and developed the electronics set-up for the CCS measurements. C.B., C.P., and Z.B. performed the spectroscopy measurements with the OPO. C.P., Z.B., and C.B. carried out the data analysis of the DCS and CCS measurements. C.B. and C.P. wrote the manuscript. C.P. and U.K. supervised the project. All authors discussed the results and read and approved the final manuscript. 	

\bmhead{Disclosures}
The authors declare no conflicts of interest.	

\bmhead{Data availability}
Data underlying the results presented in this paper is available at ETH Zurich Research Collection library~\cite{ETHZ_ResearchCollection}. 

\section*{Supplementary Information}
\subsection*{Low-pressure CCS measurement}

\begin{figure}[H] % [h]%
\centering
\includegraphics[width=0.95\textwidth]{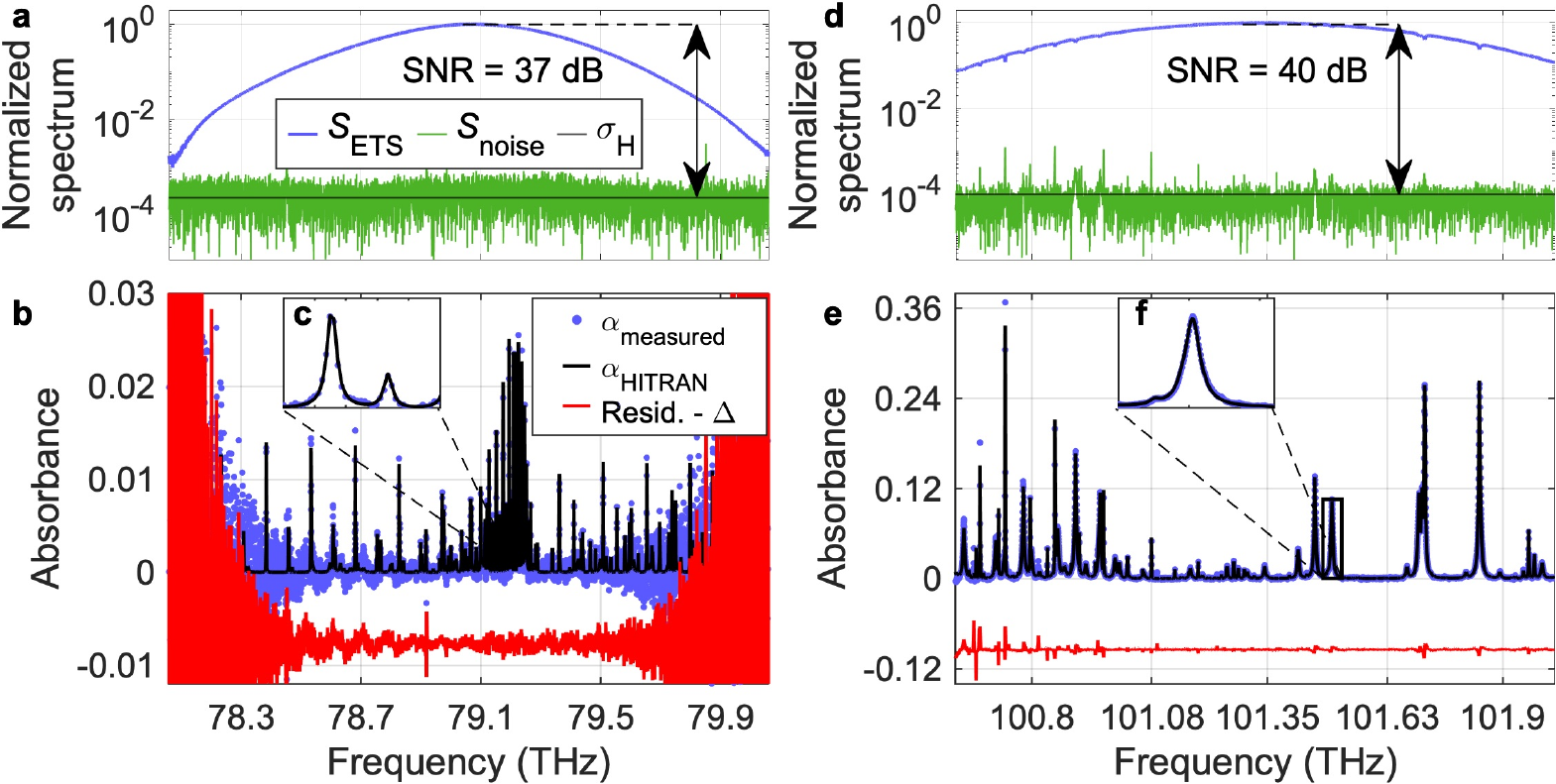}
\caption{
CCS measurements on a low-pressure acetylene cell. Left column: Measurement around 3792~\unit{\nano\meter}. Right column: Measurement around 2959~\unit{\nano\meter}. \\
\subfig{a} Measured spectrum  (blue) and its noise floor (green). The blue curve is obtained from $\etsTime{ETS}(t)$ over the time range from -20~\unit{\pico\second} to 2000~\unit{\pico\second}. The green curve is obtained from $\etsTime{noise}(t)$ over the full time range except for the small region around $t=0$ where the stitching occurs at 6~\unit{\pico\second}.\\
\subfig{b} Measured absorption spectrum (blue) compared with HITRAN calculations (black). The residuals are shown by the red curve, which is displaced from 0 for visibility. \\
\subfig{c} Zoom-in on the absorption features to highlight the good agreement of the measurement with HITRAN. $\alpha_{HITRAN}$ is calculated with a sampling rate corresponding to the resolution of the measured data.\\ 
Corresponding information for \subfig{d-f}. The blue curve is obtained from $\etsTime{ETS}(t)$ over the time range from -50~\unit{\pico\second} to 2800~\unit{\pico\second}. The stitching occurs at 20~\unit{\pico\second}.
}
\label{fig:CCS_C2H2_1420_1580}
\end{figure}

To demonstrate the suitability of the source for high-resolution measurements, we conducted two more measurements on a 65-\unit{\milli\bar} acetylene cell where the absorption features have a width $\sim 1~\unit{\giga\hertz}$. Firstly, the idler center wavelength is set to 3792~\unit{\nano\meter}, at the wings of a strong acetylene absorption feature. In this spectral band also water has some weak absorption lines. The results of this measurement are shown in Fig.~\ref{fig:CCS_C2H2_1420_1580}a-c. This measurement has a \SNRabsolute{} of 36.5~\unit{\decibel},$\sigma_H=2.2\times 10^{-4}$ for a 10-\unit{\milli\second}-long measurement, corresponding to 46.6~\dBRootHz{}, and a FOM of $1.3\times10^8~\unit{\hertz}^{1/2}$. The SNR is limited by the available idler power, which is reduced because the sample cell is not ideally suited for mid-IR measurements: it has fused silica windows which are absorbing in the wavelength range $>$2400~\unit{\nano\meter}. The measurement shows good agreement with the HITRAN database. The standard deviation of the residuals is $7.1\times 10^{-4}$ in a 1.2~\unit{\tera\hertz} wide segment centered about the peak.

In a second measurement, the idler center wavelength is set to 2959~\unit{\nano\meter}, where acetylene has a weak absorption feature. The results of this measurement are shown in Fig.~\ref{fig:CCS_C2H2_1420_1580}d-f. The measurement has a \SNRabsolute{} of 40.1~\unit{\decibel}, $\sigma_H=9.7\times 10^{-5}$, corresponding to 50.2~\dBRootHz{}, and a FOM of $3.4\times10^8~\unit{\hertz}^{1/2}$. The measurement shows good agreement with the HITRAN database, as well. The standard deviation of the residuals is $1.6 \times 10^{-3}$ in a 1.2~\unit{\tera\hertz} wide segment centered about the peak.

To obtain the best overlap with the HITRAN data base, we account for the cell temperature, deviations from the manufacturer-specified pressure, as well as an off-center beam path through the sample cell that falls within the provided tolerances. In the CCS measurements, we also account for atmospheric absorption of methane and water. A summary of the relevant parameters can be found in Tab.~\ref{tab:spectroscopy} for all spectroscopic measurements presented. The nominal values for the cell are: $L_\mathrm{nom}=14 \pm 3$~\unit{\milli\meter} and $p_\mathrm{nom}=66 \pm 13$~\unit{\milli\bar}

\begingroup
\setlength{\tabcolsep}{2pt} % Default value: 6pt
\renewcommand{\arraystretch}{1} % Default value: 1
\begin{table}[h]
\caption{Summary of the parameters for the HITRAN-based absorbance calculation for all spectroscopic measurements performed.}
\label{tab:spectroscopy}
\begin{tabular*}{\textwidth}{@{\extracolsep\fill}lcccccc}
    \toprule%
    Parameter                          & Symbol & Unit               & DCS 1540~\unit{\nano\meter} & CCS 2959~\unit{\nano\meter} & CCS 3320~\unit{\nano\meter} & CCS 3792~\unit{\nano\meter} \\
    \midrule
    Cell pressure & $p_\mathrm{cell}$     & \unit{\milli\bar} & 67.9     & 78.5   & -    & 79 \\
    Cell temperature & $T_\mathrm{cell}$  & \unit{\celsius}   & 23       & 42.9      & -    & 43.9 \\
    Cell distance & $L_\mathrm{cell}$       & \unit{\milli\meter} & 13.4   & 13.2     & -    & 13.6 \\
    Air pressure & $p_\mathrm{air}$       & \unit{\milli\bar}       & -        & $\approx$950       & $\approx$950      & $\approx$1000 \\
    
    Air temperature & $T_\mathrm{air}$    & \unit{\celsius}   & -        & 23      & 25     & 23 \\
    Air distance & $L_\mathrm{air}$         & \unit{\meter}     & -   & 2.68   & 2.71       & 2.68 \\
    Fraction $\mathrm{H}_2$O & $F_\mathrm{H_2O}$    & \%    & -     & 1.01   & 1.05        & 1.52 \\
    Fraction $\mathrm{CH}_4$ & $F_\mathrm{CH_4}$              & ppm        & -  & -     & 2.5       & - \\
    \botrule
\end{tabular*}
\end{table}
\endgroup

We attribute the elevated temperature and pressure of the acetylene cell in the CCS measurements to the strong absorption of UVFS at the idler wavelength. The fractions for water and methane are found by keeping the interaction length of the idler with the atmosphere fixed to a value found by measuring the path length with a ruler. All measurements have been conducted on different days such that variations in the air humidity can be expected. For the measurements using the acetylene cell, the optical alignment and cell placement have changed in between the measurements, such that the exact path through the cell was not preserved.

\subsection*{Coherence of the SWIR DCS measurement}

\begin{figure}[H] % [h]%
\centering
\includegraphics[width=0.9\textwidth]{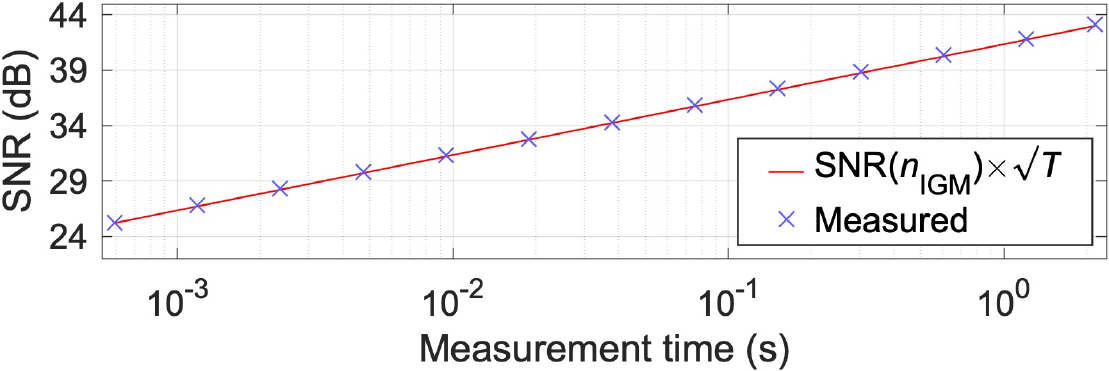}
\caption{
\SNRabsolute{} as a function of the measurement time $T$ (blue crosses), compared to the \SNRabsolute{} of a single IGM scaled with $\sqrt{T}$ (red). 
}
\label{fig:DCS_SNR}
\end{figure}

In Fig.~\ref{fig:DCS_SNR}, we show how the \SNRabsolute{} of our measurement scales by $\sigma_H\propto T^{-1/2}$~\cite{Newbury2010}, where $T$ is the measurement time.

\subsection*{Relative intensity noise measurements}

\begin{figure}[H] % [h]%
	\centering
	\includegraphics[width=0.9\textwidth]{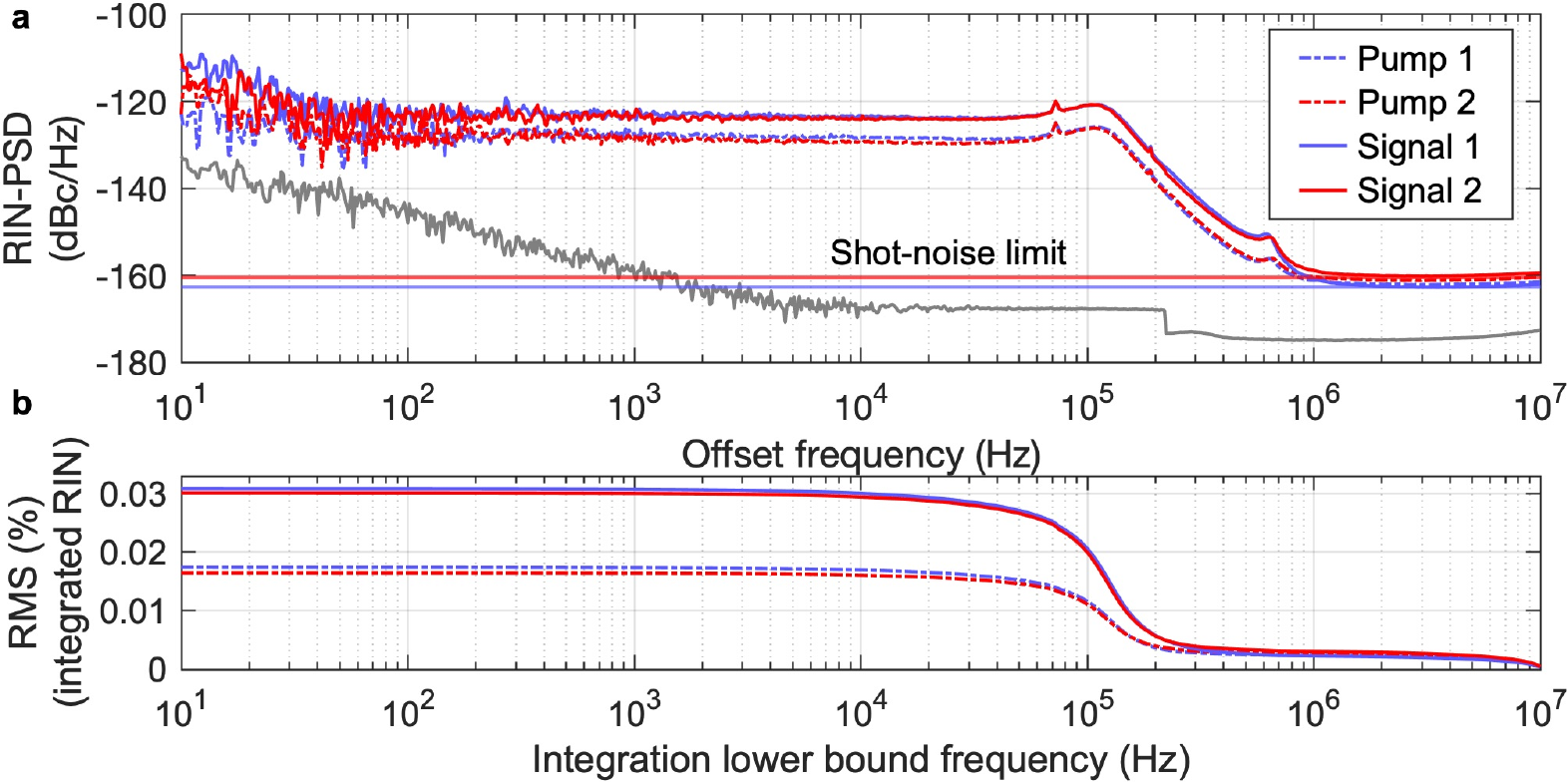}
	\caption{Intensity noise characterization of the single-cavity dual-comb system based on baseband measurements for the Yb:YAG laser (dash-dotted: pump 1 in blue, pump 2 in red) and for the signal beams (solid: signal 1 in blue, signal 2 in red). To determine the measurement noise floor, the optical input to the photodiode was blocked and the resulting power spectral density was scaled according to the typical incident power in the unblocked case (4.5~\unit{\milli\watt}). The resulting effective RIN noise floor is at around -175~\dBRootHz{} (grey); it is well below the measured RIN values. 
     \subfig{a} RIN-PSD. The straight solid lines indicate the shot noise limit for the signal measurements: -163 \dBRootHz{} (signal 1, blue), -160.5 \dBRootHz (signal 2, red). 
     \subfig{b} Integrated RIN-PSD with the integration interval from 10~\unit{\hertz} to 10~\unit{\mega\hertz}. Stitching frequency: 220~\unit{\kilo\hertz}.}
	\label{fig:RIN_results}
\end{figure}

The RIN measurement scheme, which is adapted from the one we described in~\cite{Pupeikis2022}, uses two measurements (one for low noise frequencies and one for high noise frequencies) and stitches them together to get high sensitivity across the whole frequency range. The baseband noise on the photocurrent generated by an InGaAs photodiode is measured on a signal source analyzer (SSA) (Keysight Technologies E5052B). To illustrate the relation of the pump and OPO RIN, the measurements of both the pump beams and the signal beams are given. For the RIN measurements, the following average powers were sent to the photodiode: 6.9~\unit{\milli\watt} (pump 1), 5.4~\unit{\milli\watt} (pump 2), 4.6~\unit{\milli\watt} (signal 1), and 3.0~\unit{\milli\watt} (signal 2). The RIN power spectral density (PSD) for both the SSL and the OPO spanning from 10~\unit{\hertz} to 10~\unit{\mega\hertz} is displayed in Fig.~\ref{fig:RIN_results}a, and the corresponding integrated RIN spectrum is shown in Fig.~\ref{fig:RIN_results}b. We see that the RIN reaches the measurement noise floor at about 1~\unit{\mega\hertz}. For both the pump laser and the OPO, the RIN is lower than –160~\dBRootHz{} beyond 1~\unit{\mega\hertz} offset frequencies. For the OPO, the integrated RIN is 0.031\% (signal 1) and 0.030\% (signal 2) when integrated from 10~\unit{\hertz} to 10~\unit{\mega\hertz}. Due to the absence of available data in literature, a comparison of the RIN performance to other dual-comb OPO approaches is not possible.

\bibliography{Library_4thsubmisison_OPOpaper}%

%TC:endignore
\end{document}